\documentclass[epjc3,smallextended,final,twocolumn]{svjour3}

\usepackage{cite} 
\usepackage[english]{babel} 
\usepackage[utf8]{inputenc}
\usepackage{latexsym} 
\usepackage{amsmath} 
\usepackage{amsfonts} 
\usepackage{amssymb} 
\usepackage{textcomp} 
\usepackage{graphicx}  
\usepackage{enumerate} 
\usepackage{array} 
\usepackage{mathrsfs}
\usepackage{bm,bbm}
\usepackage{bbold}  
\usepackage{multirow}
\usepackage{fancyref}
\usepackage[breaklinks]{hyperref}
\usepackage{color}
\newcommand{\Eq}[1]{(\ref{#1})}

\newcommand{\be}{\begin{equation}}
\newcommand{\ee}{\end{equation}}
\newcommand{\ba}{\begin{eqnarray}}
\newcommand{\ea}{\end{eqnarray}}
\newcommand{\bs}{\begin{subequations}}
\newcommand{\es}{\end{subequations}}
\def\com{\color{magenta}}
\def\cob{\color{blue}}
\newcommand{\rmd}{{\rm d}}

\newcommand{\arX}[1]{\href{http://arxiv.org/abs/#1}{{\ttfamily\com arXiv:#1}}}
\newcommand{\oarX}[1]{\href{http://arxiv.org/abs/#1}{{\ttfamily\com arXiv:#1}}}

\newcommand{\doin}[6]{\href{http://dx.doi.org/#1}{\cob  #2 #3 {\bf #4}, #5 (#6)}}
\newcommand{\doinn}[5]{\href{http://dx.doi.org/#1}{\cob  #2 {\bf #3}, #4 (#5)}}
\newcommand{\doij}[5]{\href{http://dx.doi.org/#1}{\cob  #2 {\bf #3}, #4 (#5)}}

\newcommand{\ndoinn}[5]{\href{#1}{\cob  #2 {\bf #3}, #4 (#5)}}
\newcommand{\book}[5]{{\it #1} (#2, #3, #5)}

\newcommand{\proc}[6]{in \emph{#1}, ed.\ by #2 (#3, #4, #6)}
\newcommand{\procsin}[5]{in \emph{#1} ed.\ by #2 (#3, #4, #5)} 
\newcommand{\tia}[1]{#1.}

\def\p{\partial}
\def\a{\alpha}
\def\b{\beta}

\def\de{\delta}
\def\g{\gamma}
\def\s{\sigma}

\def\la{\lambda}

\def\om{\omega}
\def\G{\Gamma}

\def\vr{\varrho}

\def\e{\epsilon}
\def\De{\Delta}

\def\N{\nabla}

\def\cK{\mathcal{K}}
\def\cL{\mathcal{L}}
\def\cP{\mathcal{P}}
\def\cT{\mathcal{T}}
\def\cX{\mathcal{X}}

\def\cF{\mathcal{F}}
\def\cT{\mathcal{T}}

\def\cS{\mathcal{S}}

\def\cO{\mathcal{O}}

\def\cV{\mathcal{V}}

\renewcommand{\dh}{d_\textsc{h}}
\newcommand{\ds}{d_\textsc{s}}
\newcommand{\dw}{d_\textsc{w}}

\journalname{Eur. Phys. J. C}

\date{February 3, 2016}

\begin{document}\sloppy 

\title{ABC of multi-fractal spacetimes and fractional sea turtles}

\author{Gianluca Calcagni\thanksref{addr1,e1}}
\thankstext{e1}{e-mail: calcagni@iem.cfmac.csic.es}
\institute{Instituto de Estructura de la Materia, CSIC, Serrano 121, 28006 Madrid, Spain\label{addr1}}

\maketitle

\begin{abstract}
We clarify what it means to have a spacetime fractal geometry in quantum gravity and show that its properties differ from those of usual fractals. A weak and a strong definition of multi-scale and multi-fractal spacetimes are given together with a sketch of the landscape of multi-scale theories of gravitation. Then, in the context of the fractional theory with $q$-derivatives, we explore the consequences of living in a multi-fractal spacetime. To illustrate the behavior of a non-relativistic body, we take the entertaining example of a sea turtle. We show that, when only the time direction is fractal, sea turtles swim at a faster speed than in an ordinary world, while they swim at a slower speed if only the spatial directions are fractal. The latter type of geometry is the one most commonly found in quantum gravity. For time-like fractals, relativistic objects can exceed the speed of light, but strongly so only if their size is smaller than the range of particle-physics interactions. We also find new results about log-oscillating measures, the measure presentation and their role in physical observations and in future extensions to nowhere-differentiable stochastic spacetimes. 
\end{abstract}



\section{Introduction and main results}\label{intro}

In the multi-faceted quest for a theory of quantum gravity, evidence has been gathered that quantum spacetimes acquire anomalous properties which cannot be described by conventional geometry \cite{tH93,Car09,fra1}.  Volumes and distances can change depending on their size and length, on the size of the observer, on the scale of the experiment, and so on. In particular, the dimension of spacetime changes with the scale, in a way similar to what happens in multi-fractal sets. Among the many available examples of this \emph{dimensional flow}, we count with causal dynamical triangulations \cite{AJL4,BeH,SVW1}, asymptotically safe quantum gravity \cite{LaR5,CES}, loop quantum gravity and spin foams \cite{Mod08,COT2,COT3}, Ho\v{r}ava--Lifshitz gravity \cite{SVW1,CES,Hor3}, non-commutative geometry \cite{Con06,CCM,AA} and $\kappa$-Minkowski spacetime \cite{Ben08,ACOS,ArTr1}, non-local quantum gravity \cite{Mod11}, Stelle's gravity \cite{CMNa}, spacetimes with black holes \cite{CaG,Mur12,AC1}, fuzzy spacetimes \cite{MoN}, random combs \cite{DJW1,AGW}, random multi-graphs \cite{GWZ1,GWZ2}, causal sets \cite{EiMi} and string theory \cite{CaMo1}.

Usually, the discussion of dimensional flow is maintained at a rather technical level but, when trying to translate mathematical properties into physics, it is customary to describe these geometries as ``fractal.'' Then the usual questions posed when talking about fractal spacetimes are:
\vspace{0.5cm}
\ba
&&\text{\emph{What is a fractal?}}\label{quest1a}\\
&&\text{\emph{What is a fractal spacetime?}}\label{quest1b}\\
&&\text{\emph{How would the world look like on a multi-fractal?}}\label{quest2}
\ea\vspace{0.1cm}

\noindent Sometimes, at this point one feels a slight embarrassment. Establishing a set of mathematical properties for a geometry is insufficient to acquire a complete understanding of what an observer would experience in such a geometry, especially when the ``observer'' is an abstract Planck-size probe. Moreover, the concept of fractal has been widely used in quantum gravity, but it never has received a proper definition. Part of the reason is that there is no intrinsic definition even for the popular fractals we come across in computer graphics \cite{Fal03}. As the mathematician Robert Strichartz said when asked \Eq{quest1a}, ``I know one when I see one'' \cite{Str03}. At most, we can make a list of properties we would expect the archetypical fractal should obey, marking with an asterisk optional ones:
\begin{enumerate}
\item[1.] A fine structure: the set has details at every scale.
\item[2.] An irregular structure: ordinary continuous differential calculus cannot be applied on the set.
\item[*3.] Self-similarity.
\item[*4.] A non-integer dimension (Hausdorff dimension $\dh$, spectral dimension $\ds$ or walk dimension $\dw$).
\item[5.] The relation $\dw=2\dh/\ds$ holds with $\ds\leq\dh$.
\end{enumerate}
Furthermore, a multi-fractal (a set whose fractal properties change with the scale) should have an additional feature:
\begin{enumerate}
\item[6.] Properties 1, 2 and 5 hold at any given scale in the dimensional flow.
\end{enumerate}
Notice that 6 implies that dimensional flow occurs for at least two of the dimensions $\dh$, $\ds$, and $\dw$ (otherwise, 5 would not hold at all scales).

In the first part of this paper, we revise questions \Eq{quest1a} and \Eq{quest1b} and the points of the above list in the generic context of quantum gravity and of classical spacetimes with a geometry scale hierarchy. We provide a number of arguments that address question \Eq{quest1b} and replace properties 1--6 with the following \emph{ABC of multi-scale spacetimes} (items are in order of importance):
\begin{itemize}
\item[A.] Dimensional flow occurs with three properties: [A1] At least two of the dimensions $\dh$, $\ds$, and $\dw$ vary. [A2] The flow is continuous from the infrared (IR) down to an ultraviolet (UV) cut-off (possibly trivial, in the absence of any minimal length scale). [A3] The flow occurs locally, i.e., curvature effects are ignored (this is to prevent a false positive).
\item[B.] As a byproduct of A, a non-integer dimension ($\dh$, $\ds$ or $\dw$, or all of them) is observed during dimensional flow, except at a finite number of points (e.g., the UV and the IR extrema).
\item[C.] If, in addition, the relations $\dw=2\dh/\ds$ and $\ds\leq\dh$ hold at all scales in dimensional flow, then we call the ensuing geometry a \emph{weakly multi-fractal spacetime}.
\end{itemize}
Notice that A is the definition of \emph{multi-scale spacetime}, while the more specific notion of multi-fractal spacetime appears only in the case property C holds. In the last section of the paper, we will comment on a stronger definition of multi-fractal spacetimes, which includes properties A--C plus a refinement of property 2:
\begin{itemize}
\item[D.] A geometry is a \emph{strongly multi-fractal spacetime} if, in addition of satisfying A--C, it is nowhere differentiable in the sense of integer-order derivatives, at all scales except at a finite number of points (e.g., the UV and the IR extrema).
\end{itemize}

Exploiting sea turtles as an example of non-relativistic motion, we also give an intuitive answer to question \Eq{quest2} with specific reference to the multi-fractional theory with $q$-derivatives \cite{fra4,frc1,frc2,fra7,frc7,frc10,frc11,frc12,frc13}: on a multi-fractal space, a sea turtle would travel slower at microscopic scales, while it would swim faster if time were multi-fractal. Intuitively, the structure of a fractal space would hinder the motion of the poor animal due to its irregularity (an analogy is water slowly percolating through a porous rock \cite{frc4}). On the other hand, a fractal time direction would be measured by clocks that tick more slowly than ordinary ones, so that a body would have more ordinary time to cover the same distance. Similarly, it is argued that a relativistic observer on a time-like fractal can reach a velocity superior to the speed of light. 

If the real world has an anomalous geometry, then there must exist some critical time and length scales $t_*$ and $\ell_*$ below which fractal properties begin to show up. In other words, geometry must be multi-scale or multi-fractal, as opposed to just fractal. The theory with $q$-derivative is of this form and upper bounds on $t_*$ and $\ell_*$ have been derived recently \cite{frc12,frc13}. These bounds ($\ell_* < 10^{-19}{\rm m}$, $t_* < 10^{-27}{\rm s}$) are about 20--30 orders of magnitude smaller than the scales involved in the sea-turtle thought experiment ($\ell_{\rm turtle}> 1\,{\rm m}$, $t_{\rm turtle}> 1\,{\rm s}$) and there would be no way to discriminate a turtle on such a multi-fractal from one on plain earth. For the same reason, super-luminal motion would be possible only for an object of size $\lesssim \ell_*$ in an experiment with characteristic time $\lesssim t_*$, i.e., below the scales of Standard-Model interactions. Therefore, even if the multi-fractional theory with $q$-derivatives can avoid the side effects that super-luminal travel \cite{Alc94} entails in Einstein gravity (existence of exotic matter \cite{Alc94,Eve96,VBLi} and quantum instability due to a thermal flux of Hawking particles \cite{FLiB}) or in generic Lorentz-violating set-ups \cite{CFLPa}, to curb the enthusiasm of Sci-Fi \emph{aficionados} we already anticipate that our model, or possibly any multi-fractal spacetime in general, cannot be used as a practical base for a hyperdrive.

Questions \Eq{quest1a} and \Eq{quest1b} are addressed in Sect.\ \ref{1}. Sections \ref{2} and \ref{2a} are a self-contained review of the theory with $q$-derivatives, with a new result concerning a reduction of the parameter space of measures with logarithmic oscillations. A novel discussion of the problem of presentation is given in Sect.\ \ref{2b}. Since this part of the paper is somewhat technical, we summarize its content here in a few intuitive points.
\begin{enumerate}
\item[(i)] In order to fully define a multi-fractional theory, one must choose a frame where geometric coordinates $q(x)$ are written down explicitly. This is not a return to a pre-relativistic view of space and time because the frame choice affects the integro-differential structure of the theory, not the metric structure \cite{frc11}.
\item[(ii)] There is a limited number of presentation choices; we will see four below. 
\item[(iii)] Different presentations of the same measure correspond to different theories in the same geometric class (i.e., they show the same scaling properties). In Sect.\ \ref{nodif}, this feature is restated under a new perspective based on the famous It\^{o}--Stratonovich dilemma in stochastic processes.
\item[(iv)] Although they describe the same class of geometries (iii) and they are not many (ii), different presentations may have profound consequences for physical properties such as the propagation speed of bodies or elementary particles, but only in extreme situations of high energy, high curvature or small scales.
\item[(v)] Therefore, even if the $q$-theory is invariant under Poincaré transformations on the coordinates $q$ meant as non-composite objects, the physics is not completely independent of the choice of coordinates $x$ used to describe the system in the frame where physical observables are extracted.
\end{enumerate}
The motion of a non-relativistic and a relativistic body are studied in, respectively, Sects.\ \ref{3} and \ref{4}. In Sect.\ \ref{3}, we also clarify the relation between measurement units and frame choice by noting that a discrimination between a fractal and a normal spacetime is possible when we can determine dimensionless quantities such as the ratio of two observables of the same kind. Section \ref{futu} is devoted only to open threads and future developments, including the effect of log oscillations and an interesting connection between presentation choice and stochastic processes on multi-scale spacetimes, which motivates the introduction of property D.


\section{Spacetime multi-fractals in quantum gravity}\label{1}

Let us now examine the list of properties 1--6 and how they apply to quantum gravity or, more generally, to anomalous spacetimes.


\subsection{Dimensions}

To characterize a set or a geometry, we have various operational definitions of dimension.\footnote{Mathematical statements can be found, e.g., in \cite{frc1,Fal03}.} In this subsection, we first revisit these definitions in the case of a continuum space, then commenting on theories with discrete structures and finally including also time. In the process, we will get a first glimpse of the type of phenomena we would experience if we lived in a multi-fractal world, and of how to detect them.

\subsubsection{Dimensions of continuous spaces}

In an ambient space with $D-1$ topological dimensions, the Hausdorff dimension $\dh$ is the scaling of the volume $\cV_{D-1}(\ell)$ of a $(D-1)$-ball with respect to the radius $\ell$, $\dh:=\rmd \ln\cV_{D-1}(\ell)/\rmd\ln \ell$. For a set with constant dimension,
\be\label{eqdh}
\cV_{D-1}(\ell)\sim \ell^{\dh}\,,
\ee
while for a multi-scale set $\cV_{D-1}(\ell)\propto (\ell/\ell_1)^{\dh^{(1)}}+(\ell/\ell_2)^{\dh^{(2)}}+\dots$ (the actual coefficients are a bit more involved; see \cite{frc4,fra6}). Depending on the relative size of the ball with respect to the lengths $\ell_1$, $\ell_2$,\dots, the Hausdorff dimension will be $\simeq \dh^{(1)},\dh^{(2)},\dots$. In other words, if one tries to measure the volume $\cV_{D-1}$ of a ball, its scaling with the radius is different depending on whether $\ell$ is larger or smaller than the characteristic scale at which ``fractal'' effects become apparent. An observer in a space with $\dh=D-1$ at large scales $\ell\gg\ell_*$ and $0<\dh<D-1$ at small scales $\ell\ll\ell_*$ can make several balls of radius $R_1+\de R$ close to some average value $R_1\gg \ell_*$ (where $\de R\ll \ell_*$), submerge each ball in a container of water and measure the volume of displaced liquid, noting a distribution of volumes with average $R_1^{D-1}$ and width $\sim R_1^{D-2}\de R$. Making another set of balls of average radius $\de R<R_2\ll\ell_*$ with the same fluctuation $\de R$, they find an average volume $R_2^{\dh}$ and (for $D\geq 3$ and $\dh\geq 1$) a narrower distribution, since $1\ll (R_1/\ell_*)^{D-2} > (R_2/\ell_*)^{\dh-1}\ll 1$. The inequality may change direction for $\dh<1$ but, in any case, by comparing these dimensionless observables the experimenter realizes that they are living in a space with dimensional flow.

The spectral dimension $\ds$ is the scaling of the return probability in a diffusion process. Let $\cK(\N)$ be the Laplacian of a theory on a continuum Riemannian manifold; in the standard case, $\cK(\N)=\N^2$. Placing a point-wise test particle at point $x'$ on a spatial geometry and letting it diffuse, its motion will obey the non-relativistic diffusion equation $[\p_\s-\kappa\cK(\N)]P(x,x',\s)=0$ with initial condition $P(x,x',0)=\de(x-x')/\sqrt{g}$, where $\kappa$ is a diffusion coefficient, $\s$ is an abstract diffusion time parametrizing the process and $g$ is the determinant of the metric. If $\s=t$ is the proper time of the particle or a viable global time variable, then $\kappa$ is measured in ${\rm m}^2\,{\rm s}^{-1}$. Integrating the heat kernel $P$ for coincident points over all points of the geometry, one obtains a function $\cP(\s):=Z/\cV_{D-1}=\int\rmd^{D-1}x\sqrt{g}P(x,x,\s)/\cV_{D-1}$ called return probability (the volume factor makes the normalization finite). Then $\ds:=-2\rmd\ln\cP(\s)/\rmd\ln\s$. For a set with constant spectral dimension,
\be\label{eqds}
\cP(\s)\sim \s^{-\ds/2}\,,
\ee
while for a multi-scale set the structures of $\cK$ and of a generalization of the operator $\p_\s$ determine two or more asymptotic regimes \cite{frc4}. A particle in a space with $\ds=D-1$ in the IR and $0<\ds<D-1$ in the UV diffuses slower in the ultraviolet.\footnote{In fractal geometry, the spectral dimension is also conjectured to coincide with the dimension of momentum space \cite{Akk12,Akk2}. This can easily be shown to be true in the continuum in the presence of non-trivial dispersion relations $\cK(-p^2)\neq-p^2$ \cite{AAGM3}.}

Finally, the walk dimension is the scaling of the mean-square displacement of a random walker $X(\s)$, $\dw:=2(\rmd\ln\langle X^2(\s)\rangle/\rmd\ln\s)^{-1}$, where $\langle X^2(\s)\rangle=\int\rmd^{D-1}x\,x^2\,P(x,0,\s)$. For a set with constant walk dimension,
\be\label{eqdw}
\langle X^2(\s)\rangle\sim \s^{2/\dw}\,.
\ee
In a space with walk dimension $\dw=2$ in the IR and $\dw>2$ in the UV, the ratio $(\rmd \sqrt{\langle X^2\rangle}/\rmd\s)/(\sqrt{\langle X^2\rangle}/\s) \sim 1/\dw$ between the differential velocity of the particle (measuring its local random motion) and the total finite-difference velocity decreases in the UV. This means that the trajectory of the probe becomes less ragged than usual in the UV; a pictorial demonstration of such behavior can be found in \cite{frc7}.

\subsubsection{Dimensions of discrete and combinatorial structures}

In several quantum-gravity approaches, there is no fundamental continuous spacetime. Nevertheless, it is possible to generalize the above operational definitions to a discrete set and to extract, only in certain regimes, sensible multi-scale profiles for the Hausdorff, spectral and walk dimension. A proof of concept is given in \cite{AJL4,COT2,COT3,DJW1,AGW,GWZ1,GWZ2}. For instance, a generalization of discrete exterior calculus \cite{DHLM,BeHi,COT1} allows one to construct Laplacians on combinatorial structures and hence define diffusion processes thereon. The spectral dimension as well as $\dw$ and $\dh$ can be computed for a class of quantum-gravity states, in particular those appearing in loop quantum gravity, spin foams, and group field theory \cite{COT2,COT3}.

The main requirement we ask in order to have dimensional flow in discrete (pre-)geometries is that there exist regimes where all three geometric indicators $\dh$, $\ds$, and $\dw$ are real-valued and positive. These regimes should extend from the infrared down to some effective UV scale below which discreteness or quantum effects destroy some or all of the indicators, for instance if the expectation values on the chosen states become complex. In general, the effective UV scale is determined by the choice of states. The existence of a regime where quantum geometry has well-defined dimensions translates into regularity assumptions on the quantum states \cite{COT3}. Beyond these assumptions, one can plunge into a wild jungle of quantum-geometry configurations with properties completely different from classical or semi-classical spacetimes.

\subsubsection{Dimensions of spacetimes}

In a continuous spacetime with $D$ topological dimensions, the definition of Hausdorff dimension is unchanged but for the addition of the Euclideanized time direction. Then the Hausdorff dimension of spacetime is the scaling of the volume of the $D$-ball. For the spectral dimension, one includes (imaginary) time in the operator $\cK(\N)$, while in the walk dimension time is included in Euclideanized distances $X^2(\s)=T^2(\s)+X_1^2(\s)+\cdots$. Similar considerations apply to discrete geometric or pre-geometric structures.

In the extension of all the above definitions, one takes the time direction in Euclidean signature. This step is fairly standard when one wants to define the dimension of a geometry with Lorentzian signature. If, for any reason, one cannot or does not want to Euclideanize time, then it is necessary to consider the dimensionality of spatial slices and the time line separately, instead of the whole spacetime. The reader may adopt whichever point of view they might prefer; this does not affect the following.

The spectral dimension is sometimes regarded as a theoretical parameter useful to classify spacetimes but that does not correspond to a physical observable. Elsewhere \cite{CES,CMNa}, the author and collaborators had already the occasion to advance a different view: the spectral dimension should be a meaningful observable just like the topological and Hausdorff dimension are. In that case, however, its definition must be well posed at all stages to make sense physically: if we want $\ds$ to be a physical observable (our working hypothesis here), its definition must provide also an operational way to measure it. For instance, how can we interpret the parameter $\s$ if time is in the operator $\cK$? Also, in certain cases the form of $\cK$ is such that $P(x,x',\s)$ is not a probability and there is no well-defined underlying diffusion process at all (this is a well-known problem in transport theory with higher-order or non-local operators \cite{frc4} and in quantum gravity \cite{CES,CMNa}). Mathematically there is no issue whatsoever. If $\s\neq t$, one can enact a fictitious diffusion process with some Monte-Carlo time on the geometric or pre-geometric structure one wants to explore, let it be a continuous manifold or the graph ensembles of discretized gravity. Even when $P$ is not positive semi-definite and the picture of a diffusing probe fails, to determine $\ds$ one only needs to consider closed paths and integrate over them with a certain measure. However, this is insufficient to characterize an operational way to physically measure $\ds$. In two different interpretations of the spectral dimension, valid in any regime where an effective field theory can be formulated, the diffusion equation is a renormalization-group running equations depending on the IR cut-off scale $k=f(\s)$ \cite{CES} or, alternatively, it stems from the Schwinger representation of the particle propagator and $\sqrt{\kappa\s}=\ell$ is a length scale determining the resolution $1/\ell$ at which the geometry is probed \cite{CMNa}. The interpretation of the parameter $\s$ is unimportant in the following but it is worth to mention these caveats anyway.



\subsection{Fine structure and dimensional flow}

The first property of the list 1--6 is that a (multi-)fractal set $\cF$ should have ``a fine structure.'' By this, one means that it is possible to find points of $\cF$ at all scales of observation, no matter how deeply one zooms into the set. For continuous spacetimes (among others: asymptotic safety, Ho\v{r}ava--Lifshitz gravity, non-local gravity and multi-fractional spacetimes), this requirement seems trivially satisfied and not very useful. However, a careful inspection shows that it is neither trivial nor satisfied in general. 

One of greatest Einstein's intuitions was that spacetime points do not have a physical meaning \emph{per se} unless one attaches an event to them. A spacetime devoid of particle interactions, test particles, light rays or whatever event announced by matter is an empty mathematical construct. To make sense of the idea of ``finding spacetime points at all scales,'' one should be able to concoct an experiment where the physical probe can be utilized at all scales. Of course, sometimes the same device can give us information on the physics at very different scales, as is the case with the \textsc{Planck} satellite or similar observatories of the cosmic microwave background. But, in general, we do not have a universal instrument and we need to resort to different set-ups (a telescope, a particle accelerator, \ldots) to probe the physics at different scales. 

Once having defined our ideal probe as a patchwork of instruments and experiments covering all scales of interest, the problem remains to test the spacetime structure at arbitrarily small scales. Apart from obvious technical limitations we have now and probably forever after (we cannot probe the Planck scale directly, nor energies near grand unification), it may even be theoretically impossible to reach an infinite resolution, mainly because of quantum uncertainty. This is the case of asymptotic safety, where, despite the absence of any fundamental length in the theory, a minimal length appears below which one cannot separate two points by a dynamical probe \cite{RSc1}.\footnote{Non-commutative spacetimes do have a minimal length scale, but this does not prevent them to experience dimensional flow with infinite resolution \cite{Ben08,ArTr1}.} The plethora of theories based on discrete structures is also unaffected by property 1 because there are no details below the discreteness scale. For instance, causal dynamical triangulations are a discretization of a continuum but, for any practical purpose, one cannot trust any probing of the geometry at scales comparable with the size of the triangulation cell. Loop quantum gravity and spin foams are defined on complexes that induce a minimum physical Planck-size length in the spectra of volume operators \cite{rov07}. Also, both the underlying discreteness and the combinatorial structure impose an effective UV cut-off limiting the range of scales where one can make sense of the concept of spacetime dimension, while at scales larger than the cut-off they render such dimension anomalous \cite{COT2,COT3}.

A much more important property than the fine structure is that the effective geometry must have some quantum-to-classical regime where dimensional flow takes place, otherwise one could not reach a semi-classical continuum limit where the dimension of spacetime is 4. For instance, suppose to find a dimensional flow from $\ds\sim2$ in the UV to $\ds\sim 4$ in the IR (examples of this abound in the literature \cite{tH93,Car09,fra1,AJL4,LaR5,COT3,Hor3,CaMo1,fra4}), while below the UV scale one finds a non-geometric phase where one cannot define the spectral dimension, possibly for discreteness or combinatorial effects (as in \cite{COT2,COT3}). Then below the UV scale the geometry certainly does not show a fine structure (zooming in  too much, we enter ``inside'' the building blocks of the theory, let them be lattice cells, tetrahedra or something else). During dimensional flow, the fundamental degrees of freedom (e.g., quanta of geometry, labeled complexes, and so on) group together into collective modes such that the notion of spacetime dimension makes sense and is measurable. When coarse graining the fundamental degrees of freedom, the resulting effective structure is most likely to be ``fine,'' which can be tested by finding effective dynamical equations on an effective continuum. However, this test is non-trivial and few are the cases where it can be carried on \cite{AJL4}. Usually, the only datum we know, corroborated by a numerical or analytical study of dimensional flow through all scales from the effective UV cut-off to the IR, is that discreteness effects are present but not dominant in that interval.

From this discussion, we see that theories which have dimensional flow may or may not have a fine structure at all scales. Also, theories which do not have a fine structure do not necessarily have dimensional flow (example: a canonical second-order scalar field theory on a cubic lattice), while theories which have a fine structure can describe most boring geometries (example: any canonical second-order field theory on Minkowski spacetime). We conclude that property 1 is not adequate in the context of anomalous (quantum or classical) spacetimes, many of which are \emph{not} fractal in the standard sense because they do not have a fine structure.


\subsection{Irregular structure}

An ordinary (multi-)fractal set $\cF$ has ``an irregular structure'' in the sense that it cannot be described by Euclidean geometry. A Euclidean ruler would fail to measure the total length of the Western coast of Britain \cite{Man67}. Clearly, in a physical context the geometry in the infrared must be ``regular,'' so that we should consider property 2 only at the microscopic scales of a multi-fractal spacetime. In gravitational theories, geometric probes are local and curvature effects are usually ignored when one determines the dimension of space (which would be modified by curvature even in a purely classical setting,\footnote{A classic example is the sphere $S^2$. Its surface is two-dimensional (i.e., isomorphic to a plane) only locally, while $\ds\neq 2$ at scales comparable with the curvature radius.} according to the Seeley--DeWitt formula \cite{Vas03}). However, even locally there are other effects that make the geometry non-Euclidean, for instance if gravity is quantized or in the presence of a non-trivial integro-differential structure. In the first case (quantum gravity), the collective effect of quanta of geometry is to push around the probe in an anomalous way, not experienced in a classical space. Often this induces effective operators in the dynamics, which leads to the second case (multi-fractional spacetimes). A third case consists in frameworks with an underlying discrete non-regular structure, such as the complexes found in loop quantum gravity, spin foams, and group field theory. 

All three cases can be realized in so many different ways that establishing the ``irregularity'' of a geometry is a moot point. If a spacetime shows dimensional flow locally (i.e., ignoring curvature corrections), then there must be some mechanism making it irregular. Conversely, an irregular spacetime does not have to be a (multi-)fractal unless it also has a fine structure, just like the rugged surface of a rock may not be a fractal (if we zoom in, we may discover that locally it is smooth).

In Sect.\ \ref{futu}, we will consider a more precise characterization of irregularity as one of the requirements to reproduce certain microscopic properties of stochastic processes on fractals.


\subsection{Self-similarity}

Self-similarity and self-affinity are what defines all deterministic fractals. A deterministic fractal $\cF=\bigcup_i \cS_i(\cF)$ is the union of the image of some maps $\cS_i$ which take the set $\cF$ and produce smaller copies of it (possibly deformed, if the $\cS_i$ are affinities). Not all fractals are deterministic, yet they are fractals indeed; sets with similarity ratios randomized at each iteration are of this sort and they are called random fractals. 

Since self-similarity and self-affinity are shown by a huge but non-exhaustive class of fractals, it is clear that we cannot use them to characterize spacetimes in an efficient way. The standard Poincaré transformations ${x'}^\mu=\Lambda_\nu^{\ \mu}x^\nu+a^\mu$ are affinity maps and the dynamics of a covariant field theory on Minkowski spacetime is self-affine. Yet, it is not a fractal because it has no irregular structure. On the other hand, multi-fractional spacetimes with ordinary derivatives have dimensional flow but they are neither self-similar nor self-affine \cite{fra1,frc1} and for this reason they can be used only as effective models \cite{frc11}.

Any theory of particle physics and quantum gravity worth of this name is both under analytic control and potentially predictive provided symmetries are enforced. There are no known exceptions to this rule. Whatever these symmetries are (diffeomorphism invariance, conformal invariance, supersymmetry, modular invariance, and more), they constitute a guiding principle and the backbone of the theory; they may or may not give rise to dimensional flow, which is an accidental property of geometry. This point of view is not very different from what happens when a mathematician wants to construct a fractal: first some maps are defined and then the geometry of the set is studied. However, the connection between symmetry and fractality is much more tenuous in physics and symmetry takes precedence over virtually anything else.


\subsection{Non-integer dimension}

One of the most popular features of fractals is that they have non-integer dimensions. For instance, the Hausdorff dimension $\dh$ of the middle-third Cantor set is equal to the capacity $d_\textsc{c}:=-\ln N/\ln\la=\ln 2/\ln 3\approx 0.63$. Each iteration is made of $N=2$ copies rescaled by $\la=1/3$. However, there are many fractals with integer dimension, e.g., the Mandelbrot set and its boundary (both with $\dh=2$). We refer the reader to \cite{Fal03,frc1} for definitions and more examples and counter-examples. Conversely, a set with integer dimension is a fractal only if it has an irregular structure. For instance, we can tell apart the string world-sheet from the boundary of the Mandelbrot set because there is no Virasoro algebra of operators acting on the latter \cite{CaMo1}.

On the other hand, if we have a continuous dimensional flow we expect to sample over all values of the dimension between the UV and IR terminal points, which implies that the dimension is integer only at a finite number of scales, from a minimum of one (in the infrared, where $\dh^{\rm IR}=\ds^{\rm IR}=D$ by default) to a maximum of $D+1$ (if $\dh^{\rm UV}=0$ or $\ds^{\rm UV}=0$) if dimensional flow is monotonic from the UV to the IR. (In principle, there can exist extended plateaux where the generalized dimensions have approximately constant, integer values for a continuous range of scales. However, technically such plateaux are inflection or saddle points and there is only one point therein where the dimension can be exactly integer.)


\subsection{The \texorpdfstring{$d_{\rm W}=2d_{\rm H}/d_{\rm S}>2$}{} relation}\label{dwrel}

A back-of-the-envelope argument shows the existence of a relation between the dimensions $\dh$, $\ds$, and $\dw$ of a fractal set. Let us denote by $\ell$ a length scale, be it the average displacement $\sqrt{\langle X^2\rangle}\sim\ell$ or the linear size of a volume $\cV\sim \ell^{\dh}$. We saw that $\ds$ is defined as the scaling of the return probability, Eq.\ \Eq{eqds}. The latter is a probability per unit volume, so that it scales as an inverse volume. Then
\be
\s^{-\ds/2}\stackrel{\text{\tiny\Eq{eqds}}}{\sim} \cP=\frac{Z}{\cV_{D-1}}\sim \cV_{D-1}^{-1} \stackrel{\text{\tiny\Eq{eqdh}}}{\sim} \ell^{-\dh} \stackrel{\text{\tiny\Eq{eqdw}}}{\sim} \s^{-2\dh/\dw}\,,
\ee
which implies
\be\label{dwhs}
\dw=2\frac{\dh}{\ds}\,.
\ee
This formula holds independently of the interpretation of the parameter $\s$ and the only assumption one makes is about the volume scaling of the return probability. This assumption is what characterizes fractals or non-multi-scale sets such as ordinary manifolds. Furthermore, on fractals the spectral dimension is always smaller than the Hausdorff dimension and $\dw>2$ (sub-diffusion). The specific type of sub-diffusion on fractals is called labyrinthine \cite{Sok12} because the probe is hindered by ``obstacles'' and ``dead ends'' in the geometry. Diffusion on fractals can be approximated by a diffusion equation with fractional differential operator $\p_\s^\beta$, an $x$-dependent diffusion coefficient and a friction term \cite{MeN}. We will not use the form of this diffusion equation as a criterion to define a fractal as it seems too restrictive and not as robust as the above model-independent arguments. However, we will come back to this point in Sect.\ \ref{futu}.

For multi-fractals, relation \Eq{dwhs} holds (with $\dw>2$) at any given scale.

There are various examples of multi-scale processes or geometries similar to fractals but which do not obey Eq.\ \Eq{dwhs} or for which $\dw<2$. L\'evy processes are an example well known to mathematicians (see \cite{frc4} and references therein). In the case of spacetime geometries, certain non-commutative and non-local spacetimes have a spectral dimension which grows in the UV and $\ds>\dh$ \cite{ArTr1,AC1}. Also the spacetimes of the multi-fractional theories with weighted and standard derivatives (Hermitian dual to each other) are not multi-fractals, since $\dw=2D/\ds$ and $\dh\neq D$ \cite{frc7}. On the other hand, the theory with $q$-derivatives (discussed below) respects Eq.~\Eq{dwhs} \cite{frc7} and so does asymptotically safe quantum gravity (where $\dh=D$) \cite{ReS11} (see also \cite{CES}). We do not have data about the walk dimension in the other approaches mentioned in the introduction.


\subsection{ABC of spacetime fractals}

From what seen in this section, the most general and powerful characterization of anomalous spacetimes is dimensional flow. In theoretical physics, having a fine or irregular structure is not so much important as having a set of fundamental symmetries in action, but the connection between symmetries and dimensional flow is rarely immediate (exceptions are multi-fractional theories). Also, relation \Eq{dwhs} is not obeyed in several cases.

Since all the main properties of fractals are either violated or modified in quantum gravity and spacetime theories with anomalous geometry, it is highly recommended to shift the attention to the more practical notion of multi-scale spacetimes, defined with property A (and the ancillary feature B) in the introduction. If Eq.~\Eq{dwhs} and $\ds\leq\dh$ hold, then a multi-scale spacetime is also multi-fractal (property C).

\begin{figure}
\centering
\includegraphics[width=8cm]{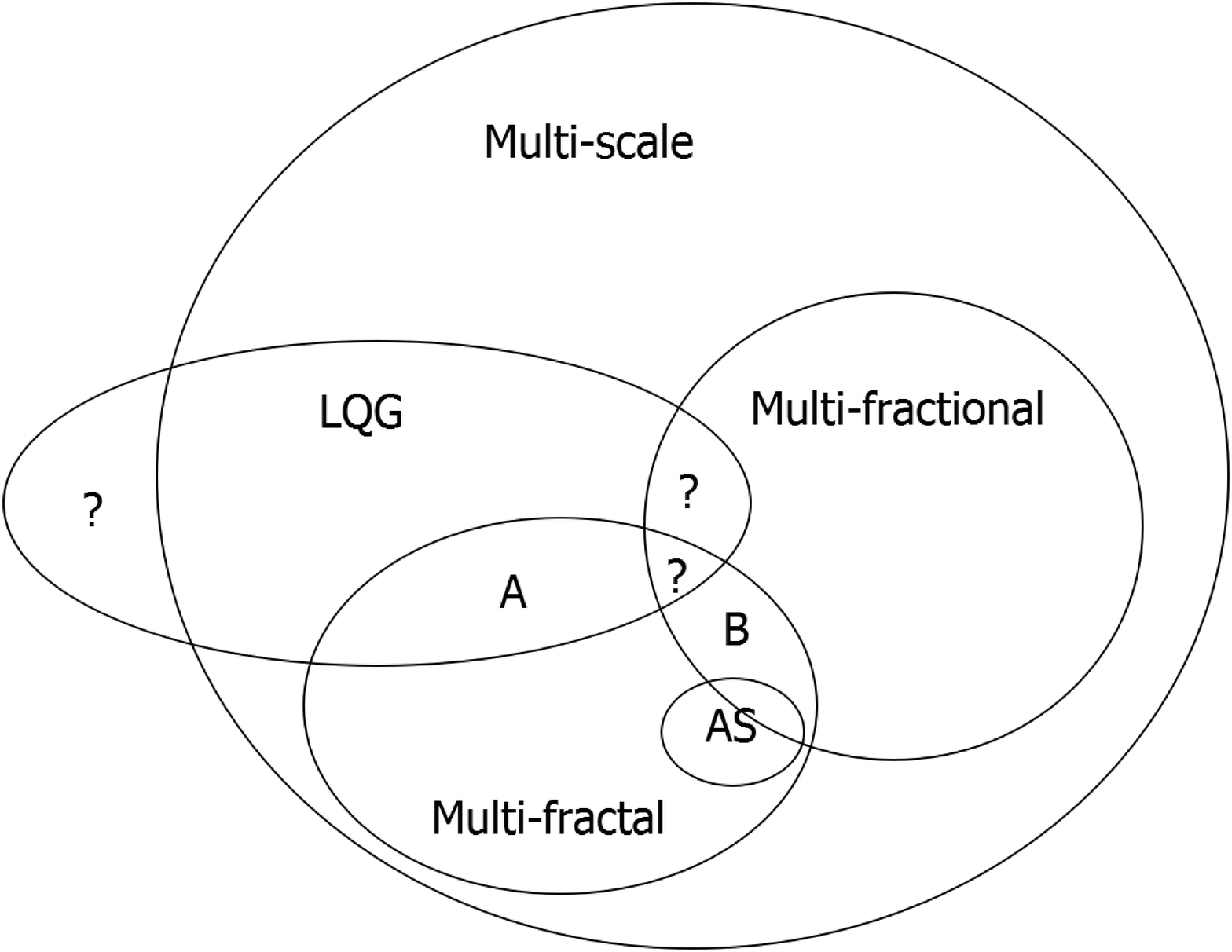}
\caption{The landscape of multi-scale theories with anomalous geometry.}
\label{fig1}
\end{figure}

In Fig.\ \ref{fig1}, we depict a first snapshot of the landscape of multi-scale theories with dimensional flow. Apart from the sub-class of multi-fractal spacetimes, some quantum-gravity frameworks are also indicated: asymptotically safe quantum gravity (AS) and loop quantum gravity (LQG). While AS realizes multi-fractal spacetimes in all known cases (see, e.g., \cite{CES}; but the situation may change as the framework evolves), the type of geometry produced by LQG depends on the states chosen in the expectation values of the operators defining the dimensions. In \cite{COT3}, one can see several examples of states corresponding to a multi-fractal geometry (region A), to multi-scale but not multi-fractal quantum geometries, and to highly quantum geometries which cannot be classified by conventional geometric indicators. The latter case is the corner ``?''\! lying outside the multi-scale landscape. A third class of scenarios is the one of \emph{multi-fractional spacetimes}, which are not necessarily of quantum gravity.\footnote{Previously in the literature, this class was often dubbed ``multi-scale'' but, after clarifying the nomenclature, it is better to stick with the name multi-fractional, leaving the term multi-scale to a much wider landscape of theories.} Of the four multi-fractional models proposed (with ordinary, weighted, $q$- and fractional derivatives), two do not realize multi-fractal spacetimes (theories with ordinary and weighted derivatives), one has not been analyzed in full detail yet (theory with fractional derivatives) and the fourth lives on multi-fractal spacetimes. Region B includes the theory with $q$-derivatives and probably also the one with fractional derivatives.

The theory with $q$-derivatives is useful to describe the renormalization-group flow of asymptotic safety in an alternative way \cite{fra7}; this is represented by an overlap between the AS set and the multi-fractional one. Used as effective descriptions of geometry, multi-fractional models can reproduce the dimensional flow of other theories. This connection has not been shown for LQG and is indicated here with the intersections ``?'' inside the landscape. Many other well-studied multi-scale theories are not shown either (including non-commutative spacetimes and Ho\v{r}ava--Lifshitz gravity, both of which do have an overlap with multi-fractional models \cite{ACOS,fra7}, and dynamical triangulations), because the walk dimension has not been calculated yet and we are presently unable to verify that $\dw=2\dh/\ds$. However, the overwhelming majority of quantum-gravity cases have $\ds\leq\dh$, which would put them inside the multi-fractal region if property C were confirmed.

In the following, we will concentrate on the multi-fractional theory with $q$-derivatives to illustrate what we could expect to see in a multi-fractal spacetime.

\section{Multi-fractional theory with \texorpdfstring{$q$}{}-derivatives}


\subsection{Sketch}\label{2}

Multi-fractional theories are realizations of anomalous geometries rather than frameworks for quantum gravity, but they can be used also as models describing effective regimes of other proposals and, as in this paper (where gravity does not play any role), to clarify what we mean by fractal spacetimes.

We begin with a brief review of the multi-fractional theory with $q$-derivatives; more information can be found in \cite{frc11,frc13}. In $D$ topological dimensions ($D=1+3$ in our case), the dynamics is defined with respect to a geometry endowed with characteristic scales. In practice, one takes their favorite action $S[1,\p_x,\phi^i]=\int\rmd^Dx\,\cL[\p_x\phi^i,\phi^i]$ (here the first entry in the left-hand side is the measure weight in the action, $\phi^i$ are some generic degrees of freedom and we are ignoring gravity) and makes the replacement $x^\mu\to q^\mu(x^\mu)$ everywhere (including the derivatives, $\p_x\to \p_{q(x)}$, hence the name of the theory). The profiles $q^\mu(x^\mu)$ are called geometric coordinates and the theory is invariant under the non-linear $q$-Poincaré transformations
\be\label{qlort} 
{q}^\mu({x'}^\mu)=\Lambda_\nu^{\ \mu}q^\nu(x^\nu)+a^\mu\,,
\ee
where $\Lambda_\nu^{\ \mu}$ are the usual Lorentz matrices and $a^\mu$ is a constant vector. The system can be written in two different ways,
\be\label{intfra}
S[1,\p_q,\phi^i]=S[v,v^{-1}\p_x,\phi^i]\,,
\ee
where $v=\det|\rmd q^\mu/\rmd x^\mu|$. The left-hand side is the frame described by the geometric coordinates $q$, called the integer picture, and it is the starting point to formulate the theory. The right-hand side is the frame, called the fractional picture, where the $x$-dependence of the geometric coordinates $q(x)$ is manifest. 

To complete the definition of the theory, we still need two data: the choice of profiles $q(x)$ and the choice of frame. If we want our continuous spacetime to change dimension with the scale, we must be able to tell the difference between ``large'' and ``small'' distances, or between ``early'' and ``late'' times. For this purpose, we can introduce at least one characteristic length $\ell_*$ and one characteristic time $t_*$ in the choice of $q(x)$ (a more general hierarchy is discussed in \cite{fra4,frc2}). It turns out that, in $D=1$ dimension, dimensional flow can be achieved by exactly the same type of measure $q(x)\sim |x|^\a$ (called fractional measure, where $0<\a<1$) one would obtain if one approximated a fractal dust on a continuum line \cite{frc2}. To get a \emph{multi}-fractal, it is sufficient to add several power-law contributions $|x|^{\a_l}$ with different $\a_l$, each multiplied by a characteristic scale $\ell_*^{(l)}$. 

For a binomial measure (just two terms, one scale $\ell_*$) and requiring to have a standard geometry in the infrared ($\alpha_1=1$, $\alpha_2=\alpha$), we have
\be\label{qx}
q(x) = q_*(x) := x+\frac{\ell_*}{\a}\left|\frac{x}{\ell_*}\right|^\a.
\ee
Generalizing to $D-1$ spatial directions, one takes the Cartesian product of $D-1$ anomalous lines and the multi-fractional spatial measure $\rmd q_*(x^1)\,\rmd q_*(x^2)\ldots\rmd q_*(x^{D-1})$, possibly with different exponents $\a_i$ \cite{frc2}. This \emph{Ansatz} for a factorizable measure is not only sufficient for our aims but it may also be necessary for technical reasons explained elsewhere \cite{frc5}. To complete the definition of the measure, we take a copy of \Eq{qx} also in time,
\be\label{qt}
q(t) = q_*(t) := t+\frac{t_*}{\a_0}\left|\frac{t}{t_*}\right|^{\a_0}.
\ee
In the following, we will consider an isotropic configuration where all spatial $\a_i=\a$ and $\ell_*^i=\ell_*$ are the same, so that the parameters of the theory in position space are $\ell_*$, $t_*$ and the two fractional exponents $\a_0$ and $ \a$. 

In momentum space, we must define geometric coordinates according to the law
\be\label{pmuk}
p^\mu(k^\mu)=\frac{1}{q^\mu(1/k^\mu)}\,,
\ee
where $\ell_*\to 1/k_*$ and $t_*\to 1/E_*$ \cite{frc11,frc13}. We will denote as $p_*(k^\mu)$ the measure dual to the binomial measure \Eq{qx}--\Eq{qt}.

Having chosen the profiles $q^\mu(x^\mu)$, let us consider the choice of frame. We must select in which picture physical observables are computed. On an ordinary manifold, the properties of clocks, rods and detectors are the same independently of the scale at which measurements are taken. In contrast, multi-fractional spacetimes are a framework where physical measurements are performed with instruments which do not adapt with the observation scale even if the geometry does \cite{fra7}. This adaptation is encoded in the structure of the fractional coordinates (i.e., of the integration measure and of differential operators), where characteristic time, length and energy scales appear.

Specifying units for the coordinates clarifies the point. In $c=1$ units, time and spatial coordinates scale as $[x^\mu]=-1$ and so do the characteristic scales ($[\ell_*]=-1=[t_*]$) and the geometric coordinates ($[q^\mu(x^\mu)]=-1$). However, in the ultraviolet the variable dependence of $q^\mu$ has an anomalous scaling $\simeq[|x^\mu|^{\a_\mu}]= -\a_\mu$, which implies that $q$-clocks and $q$-rods adapt with the scale of the experiment. Since our actual clocks and rods are non-adaptive rigid apparatus, observables should be compared with experiments in the fractional picture. A more detailed discussion can be found in \cite{frc13} but in Sect.\ \ref{3} we will make an important observation so far overlooked in the literature. One may be confused by the above argument relying on the anomalous scaling of the variable part of $q$. However, even if $x$ and $q$ have the same measurement units $[x]=-1=[q]$ exactly, it is possible to recognize a standard spacetime from an anomalous one by measuring dimensionless quantities such as ratios of observables.


\subsection{Multi-fractal properties}\label{2a}

We now show, by recalling prior results and finding new ones, how the theory knows the ABC of spacetime fractals. We will do this by a heuristic computation of the dimensions for a Euclideanized no-scale geometry with isotropic profiles $q^\mu\propto |x^\mu|^\a$, where $\a_0=\a$. (The dimension of space is obtained by ignoring the time contribution.) Exact results and the multi-scale case can be found in \cite{frc2,frc4,frc7}. The Hausdorff dimension of a $D$-ball with radius $\ell$ centered at the origin is $\cV_{D}(\ell)\propto L^D(\ell)$, where $L^2(\ell)=\sum_\mu [q^\mu(\ell)]^2$. From that, one finds $\dh=D\a$. The spectral dimension descends from the diffusion equation
\be\label{pqb}
\left[\p_{q_\b(\s)}-\kappa\N_{q(x)}^2\right]P=0\,,
\ee
where $q_\b\propto\s^\b$, and reads $\ds=-2\rmd\ln\cP/\rmd\ln\s=-2\b\rmd\ln\cP/\rmd\ln q_\b(\s)=D\b$. The walk dimension $\dw=\a/\b$ is read from $\langle X^2\rangle\propto q_\b^{1/\a}$. Then Eq.\ \Eq{dwhs} holds for any $\b$. If $\b\leq \a$, then $\ds\leq\dh$ and the geometry is fractal; one has equality if $\s$ is a genuine diffusion time or if one fixes the ambiguity parameter $\b$ to be the average fractional charge $\a$ \cite{frc7}.\footnote{The anisotropic case where some or all $\a_\mu$ are different is tricky because the components of the random walk $X^2=X_0^2+X_1^2+\cdots$ would have inhomogeneous scaling and each direction should be considered separately. This complication is responsible of the fact that the dimension of the Cartesian product of fractals may not coincide with the sum of their dimensions \cite{Fal03}.}

In Sect.\ \ref{dwrel}, we mentioned that the effective diffusion equation on fractals is modified not only in the spatial part (Laplacian, diffusion coefficient, friction terms) but also in the time part, via a fractional diffusion operator. Although we have not included these modifications as part of the definition of fractals (to the best of our knowledge, this type of diffusion equation is only an empirical tool to describe transport on fractal media), they teach us that sub-diffusion on a space-like fractal may come from a diffusive process parametrized by an anomalous clock $\s$. From what we know about diffusion in multi-scale spacetimes with $q$-derivatives \cite{frc7}, we recognize a diffusion operator $\p_{q_\b(\s)}$ in Eq.\ \Eq{pqb} which is not fractional but it is anomalous nevertheless. Moreover, expanding $\kappa \N_{q(x)}^2$ in $x$ coordinates we find both an effective space-dependent diffusion coefficient $\sim\kappa/(\p_x q)^2$ and a first-order friction term. All the ingredients of fractal diffusion are here, albeit modified with respect to the phenomenological models of \cite{MeN}.

The geometry of the $q$-theory is a random fractal, namely, a fractal endowed with symmetries whose parameters are randomized each time they are applied over the set \cite{frc2,RLWQ}. To get a deterministic fractal where the symmetry parameters are fixed (the Cantor sets and the Koch curve are examples), it is sufficient to include logarithmic oscillations of the coordinates in Eqs.\ \Eq{qx} and \Eq{qt} \cite{frc2,NLM}. Then, for each direction and in dimensionless units, one replaces $q(x)=\sum_l q_{\a_l}(x)$ with $q_{\rm log}(x)=\sum_l q_{\a_l}(x) F_{\om_l}(\ln|x|)$, where $\om_l$ is a frequency parameter and $F_{\om_l}=1+A\cos(\om_l\ln|x|)+B\sin(\om_l\ln|x|)$. The fractal $\cF=\otimes_\mu \cF_\mu$ represented by a measure with only one frequency $\om>0$ is given, for each direction, by the union of $N$ copies of itself rescaled by a factor $\la_\om=\exp(-2\pi/\om)$ at each iteration. Since the capacity of $\cF_\mu$ is equal to the Hausdorff dimension and reads $d_\textsc{c}=-\ln N/\ln\la_\om=\dh=\a$, the number of copies is
\be
N=\exp(-\a\ln\la_\om)=\exp\left(\frac{2\pi\a}{\om}\right)\,.
\ee
This formula is implicit in the results of \cite{frc2} but here we recognize a new element that shrinks the parameter space of the theory considerably: since $N$ is a positive integer, then $\om$ can only take the irrational values
\be\label{omN}
\om=\om_N:=\frac{2\pi\a}{\ln N}\,.
\ee
For $\a=1/2$ and $N=2,3,\ldots$, we have $\la_\om=1/N^2$ and
\ba
&& N=2\,,\qquad \om_2\approx 4.53\,,\qquad \la_\om=\tfrac14\,,\nonumber\\
&& N=3\,,\qquad \om_3\approx 2.86\,,\qquad \la_\om=\tfrac19\,,\nonumber\\
&& \vdots\nonumber\\
&& N=10\,,\qquad \om_{10}\approx 1.36\,,\qquad \la_\om=\tfrac{1}{100}\,.\nonumber\\
&& \vdots\nonumber
\ea
The case $N=1$ is not a fractal [Eq.\ \Eq{omN} is ill defined then], while for each $N$ one has a different fractal in the same class. To understand this point, one can take the similar case of Cantor dusts on the interval $[0,1]$. These are sets that differ from one another only by the values of the parameters $\la_{1,2}$ and $a$ in the similarity maps $\cS_1(x)=\la_1 x$ and $\cS_2(x)=\la_2 x+a$; the middle-third (or ternary) Cantor set is only one member of the class, with $\la_1=\la_2=1/3$ and $a=2/3$.


\subsection{The problem of presentation}\label{2b}

In this subsection, we analyze the effect of changes in the presentation of the geometric coordinates $q^\mu(x^\mu)$.

By construction, the symmetries of the system are the $q$-Poincaré transformations \Eq{qlort}, not the ordinary ones, and the laws of physics are invariant accordingly. However, physical observables are determined in the fractional frame and, therefore, they are not invariant under these transformations. Suppose one wishes to measure the distance $\De x$ of two points A and B in a sheet of paper. If the paper is charted by a Cartesian system, then the distance is given by the Euclidean norm
\be\label{xba}
\De x:=\sqrt{|x_{\rm B}^1-x_{\rm A}^1|^2+|x_{\rm B}^2-x_{\rm A}^2|^2}\,.
\ee
Then we make a coordinate transformation $x^i\to {x'}^i$ such that $\De x=F({x_{\rm A}'}^i,{x_{\rm B}'}^i)$ is a function of the new coordinates. For instance, going to polar coordinates $\{x^1,x^2\}\to \{\vr,\theta\}$ conveniently centered at $x_A$, one has $\De x=r$. The observed value of the distance is insensitive to the coordinates we choose to represent $\De x$ with.

In the theory with $q$-derivatives, we repeat exactly the same discussion in the fractional picture, which is one of the coordinate frames $\{x\}$ where the distance $\De x$ is calculated. However, to each of these fractional frames we must associate an integer frame described by geometric coordinates. Thus, the Cartesian fractional frame $\{x^1,x^2\}$ is mapped into the integer frame $\{q^1(x^1),q^2(x^2)\}$ and, after inverting to $x^i=x^i(q^i)$ (assuming it possible, which is not always the case) the Euclidean norm \Eq{xba} is mapped into some complicated expression $\De x(q_{\rm A}^i,q_{\rm B}^i)$ which differs from the geometric Euclidean norm $\De q:=\sqrt{\sum_{i=1}^2|q_{\rm B}^i-q_{\rm A}^i|^2}$. Below we will calculate the difference and encode it in functions $\cX^i$. But now we redo the mapping to geometric coordinates starting from polar fractional coordinates. The new integer frame is $\{q_r(r),q_\theta(\theta)\}$, where the relations between $q_r$ and the $q^i$ are $q^1=q_r\cos q_\theta$ and $q^2=q_r\sin q_\theta$.

Having recalled the rather self-evident fact that arbitrary changes of chart $\{x^\mu\}\to \{{x'}^\mu\}$ modify $q(x)$, the question is: On which chart are Eqs.\ \Eq{qx} and \Eq{qt} represented? In the example of the paper sheet, is Eq.~\Eq{qx} the form of $q$ in the integer frame $\{q^1(x^1),q^2(x^2)\}$ based on Cartesian coordinates $\{x^1,x^2\}$ or the form of $q$ in the integer frame $\{q^1(r),q^2(\theta)\}$ based on polar coordinates $\{r,\theta\}$ (so that $q^1(r)=r+(\ell_*/\a)(r/\ell_*)^{\a}$), or something else? Ordinary Poincar\'e invariance is violated by factorizable multi-scale measures. A change of presentation such as a translation, a rotation of the coordinates or an ordinary Lorentz transformation modify the size of the multi-scale corrections $\cX$ and $\cT$ defined below and one realizes that different choices of the fractional frame lead to a different theory in the integer frame. Clearly, $q^1(r)\neq \sqrt{[q^1(x^1)]^2+[q^2(x^2)]^2}$ due to the non-linear terms in expressions such as Eq.\ \Eq{qx}.

Starting from \cite{frc2}, the tacit assumption has been that Eqs.\ \Eq{qx} and \Eq{qt} are based upon the Minkowski frame where all coordinate axes are orthogonal. So far this assumption has not been discussed in detail. We fill this gap here.

First and foremost, a change of presentation changes the theory (i.e., the magnitude of the corrections $\cX$ and $\cT$) but not its qualitative features. It is well known that inequivalent presentations leave the anomalous scaling of the measure and the dimension of spacetime untouched \cite{frc1,frc2}. Therefore, multi-fractional scenarios are robust across different presentations. Picking a presentation allows us to make predictions which will change in another presentation, but not by much.

Second, the choice of the Cartesian or Minkowski fractional frame is not so restrictive as it might seem. In the physical examples studied in the literature, the observations studied in the theory with $q$-derivatives involved: (a) the decay rate of the muon \cite{frc12,frc13}; (b) the Lamb shift in the spectrum of hydrogenic atoms \cite{frc12,frc13}; (c) the cosmic-microwave-background black-body spectrum \cite{frc14}; (d) the cosmic-microwave-background temperature spectrum \cite{frc14}. In the theory with weighted derivatives, we studied (b$^\prime$)$=$(b), (c$^\prime$)$=$(c) and (e$^\prime$) the fine-structure constant determined from the light of quasars \cite{frc8}. In (a), the multi-scale correction is only time dependent and $t$ is the muon lifetime. In (b), (c), and (c$^\prime$) the multi-scale correction is energy- or temperature-dependent and, as we will argue below, this poses no problem of presentation. In (d), the spectrum is written as a function of the absolute value $|{\bf k}|$ of comoving spatial momentum; since we use Cartesian momenta $\{p^1(k^1),p^2(k^2),p^3(k^3)\}$, the expression in the fractional frame is in terms of $\tilde k:=\sqrt{[p^1(k^1)]^2+[p^2(k^2)]^2+[p^3(k^3)]^2}=|{\bf k}|+\cdots$ but, to \emph{leading} order in the multi-scale correction, it is not different from what one would have obtained using a profile $p(|{\bf k}|)$. In (b$^\prime$), the multi-scale correction depends on the characteristic time $t$ of the electromagnetic processes involved in the Lamb shift. In (e$^\prime$), the correction depends on the cosmic time $t$ of emission of light of distant objects since the big bang.
 All these settings are characterized by an effective one-dimensional multi-scale correction, either of rest-frame energies or of a well-defined time variable. There is not much arbitrariness here and the Cartesian or Minkowski chart fits the purpose. 

However, there is one last bit of ambiguity which deserves our attention: a translation
\be\label{prese}
q(x)\to \bar q(x)=q(x-\bar x)\,.
\ee
Consider one dimension, the spatial interval $\De x=|x_{\rm B}-x_{\rm A}|$ between two points A and B and its geometric analog for a binomial measure:
\ba
\De \bar q_*(x) &=& |\bar q(x_{\rm B})-\bar q(x_{\rm A})|\nonumber\\
 &=& \left|(x_{\rm B}-\bar x)+\frac{\ell_*}{\a}\left|\frac{x_{\rm B}-\bar x}{\ell_*}\right|^{\a}\right.\nonumber\\
&&\left.-(x_{\rm A}-\bar x)-\frac{\ell_*}{\a}\left|\frac{x_{\rm A}-\bar x}{\ell_*}\right|^{\a}\right|\nonumber\\
&=:&\De x|1+\cX|\,.\label{Dex}
\ea
Extending this result to time intervals $\De t:=|t_{\rm B}-t_{\rm A}|$, one has a similar expression:
\bs\ba
\De \bar q_*(t) &=&\De t|1+\cT|\,,\\
\cT &:=&\frac{1}{\a_0}\frac{t_*}{\De t}\left(\left|\frac{t_{\rm B}-\bar t}{t_*}\right|^{\a_0}-\left|\frac{t_{\rm A}-\bar t}{t_*}\right|^{\a_0}\right).
\ea\es
If we do not fix the presentation (i.e., $\bar x$ and $\bar t$), we place ourselves in a quandary. Take for definiteness $x_{\rm A}=100\,{\rm m}$ and $x_{\rm B}=200\,{\rm m}$. Then, for $\a=1/2$ (realistically found in the theory) and $\ell_*=1\,{\rm mm}$ (unrealistically large), the constraint $|\cX|\ll 1$ holds for any presentation (consistently with $\ell_*/\De x\ll 1$; this is the universal behavior of different presentations announced above) but the actual value and sign of $\cX$ change. Conversely, fixing $x_{\rm B}-x_{\rm A}$ and $\bar x$ to some random numbers while varying with respect to $x_{\rm A}$ leads to a similar spectrum of values of both signs. A multi-fractional theory without a prescription on the measure presentation is not defined completely and, therefore, would not be predictive even if we knew the scales $\ell_*$ and $t_*$.

We will call \emph{null presentation} the one with $\bar x=0$ and $\bar t=0$ and denote by
\bs\label{nullX}\ba
\cX&=&\cX_0:=\frac{1}{\a}\frac{\ell_*}{\De x}\left(\left|\frac{x_{\rm B}}{\ell_*}\right|^{\a}-\left|\frac{x_{\rm A}}{\ell_*}\right|^{\a}\right),\\
\cT&=&\cT_0:=\frac{1}{\a_0}\frac{t_*}{\De t}\left(\left|\frac{t_{\rm B}}{t_*}\right|^{\a_0}-\left|\frac{t_{\rm A}}{t_*}\right|^{\a_0}\right),
\ea\es
the associated multi-scale corrections. This is the choice made in most of the previous papers. Among all other possible presentations, there are three such that $\cX(x_{\rm A},x_{\rm B})=\cX(x_{\rm A}-x_{\rm B})$ and, therefore, do not depend on translations of the coordinate frame. In \cite{frc12,frc13}, we made the natural identification of $\bar x$ with the starting point of the experiment, in this case $x_{\rm A}$. Setting instead $\bar x=x_{\rm B}$ would lead to exactly the same result but with $\cX\to -\cX$. Let us call these prescriptions \emph{initial-point presentation} and \emph{final-point presentation}, respectively. The most uninteresting case is the \emph{symmetrized presentation} $\bar x=(x_{\rm B}+x_{\rm A})/2$, which yields $\cX\equiv 0$ and a trivial theory. 

Calling $q_\pm$ the geometric coordinates in the initial- and final-point presentation, we obtain the expressions $\De q_\pm(x)=\De x|1+\cX_\pm|$ and $\De q_\pm(t)=\De t|1+\cT_\pm|$, where
\bs\label{XT}\ba
\cX&=&\cX_\pm:=\pm\frac{1}{\a}\left|\frac{\ell_*}{\De x}\right|^{1-\a},\\
\cT&=&\cT_\pm:=\pm\frac{1}{\a_0}\left|\frac{t_*}{\De t}\right|^{1-\a_0}.
\ea\es
The sign depends on the choice between initial-point presentation ($+$) and final-point presentation ($-$). In $D$ dimensions, for each spatial direction $x^i$ one has a copy of $\cX^i=\cX(x^i)$.

Let us now discuss the problem of presentation in Fourier space and consider the momentum \Eq{pmuk} dual to the binomial geometric coordinate $\bar q_*(t)$. Including also an arbitrary constant $\bar E$, one has the energy measure
\be\nonumber
\bar p_*(E)=\left[\frac{1}{E-\bar E}+\frac{{\rm sgn}(E-\bar E)}{E_*\a_0}\left|\frac{E_*}{E-\bar E}\right|^{\a_0}\right]^{-1}.
\ee
The energy in this equation is simply $E=k^0$, the time component of the $D$-momentum, and can take either positive or negative values. The parameter $\bar E$ can be identified with the energy of the ground state of the system. The natural choice in quantum field theory is $\bar E=0$ \cite{frc12,frc13}, but in certain phenomenological situations one might want to measure the energy with respect to the ground state. For instance, a constraint on the fundamental energy scale $E_*$ was found in \cite{frc12} by asking that effects of the multi-fractal geometry be smaller than the experimental error $\de E$ on the $2S-2P$ Lamb shift $\Delta E$ in the hydrogen atom:
\be\label{pfin2}
E_* >\left(\frac{\a_0}{2-\a_0}\frac{\delta E}{\Delta E}\right)^{\frac{1}{\a_0-1}}|E_{2S}-\bar E|\,.
\ee
When $\bar E=0$, this lower limit is $E_*> 35\,\text{MeV}$ for generic $\a_0$ and $E_*> 450\,\text{GeV}$ for $\a_0=1/2$. If we choose instead $\bar E=E_{1S}$, the last factor in Eq.\ \Eq{pfin2} changes from $E_{2S}\approx-3.4\,\text{eV}$ to $E_{2S}\to E_{2S}-E_{1S}\approx 10.4\,\text{eV}$ and the bound increases about three times. From this example, we can imagine that, in general, only a positive detection of multi-scale effects would be able to rule out one presentation instead of another. This expectation is confirmed by the following analysis.

 
\section{Fractional sea turtles}\label{3}

To illustrate the effects of both the presentation choice and the picture selection (i.e., the problem of using non-adaptive instruments in a scale-dependent environment), let us consider a $(1+1)$-dimensional non-relativistic experiment where a wildlife ranger wants to check whether they live in a smooth Minkowski spacetime or, more interestingly, in a multi-scale spacetime with binomial measure $q_*(x-\bar x)$. To this purpose, they observe an adult sea turtle of size $L \sim 1\,{\rm m}$ that, after laying her eggs on a beach, enters the waters at point $x_{\rm A}$ at time $t_{\rm A}$ (event A) and reaches a buoy at point $x_{\rm B}$ at a later time $t_{\rm B}$ (event B). The ranger knows that, according to the theory, measurements are performed in the fractional picture, which is the right-hand side of Eq.~\Eq{intfra}. Here, coordinates $x$ are non-composite (i.e., their scaling is one and the same at all scales) and the action $S=\int\rmd^2x\,v(x)\,\cL$ of the system has a non-trivial measure weight $v(x)=\det|\rmd q^\mu/\rmd x^\mu|$ breaking Lorentz invariance and deforming kinetic terms. The observer will use clocks and rods which do not adapt with the scale of the observed object, a standard analog wrist watch measuring time intervals $\De t$ and a rigid rod measuring distances $\De x$ in metric units. Recall that coordinates are ``adaptive'' or ``non-adaptive'' depending on whether they are composite objects or not.

We call the ranger consulting the ``fractional $t$-watch'' a \emph{fractional observer} $\cO_x$, to distinguish them from an \emph{integer observer} $\cO_q$ which would use an adaptive ``integer $q$-watch'' measuring intervals $\De q_*(t)$. For $\cO_q$, spacetime is ordinary Minkowski when expressed in terms of the composite (adaptive) coordinates $q$ and the action is $S=\int\rmd^2 q\,\cL$. In other words, an integer observer is an ordinary observer in an ordinary, non-multi-fractal world.

Below, we omit the subscript * and use the symbols $v_x$ and $v_q$ to indicate the velocities
\be
v_x:=\frac{\De x}{\De t}\,,\qquad v_q:=\frac{\De q(x)}{\De q(t)}\,,
\ee
not to be confused with the measure weight $v=\p_x q$ of the rest of the literature.


\subsection{Preliminaries}

The ranger notices that their $t$-watch ticks 20 times between the events A and B, so that $\Delta t= |t_{\rm B}-t_{\rm A}|=20\,{\rm s}$. On the other hand, the geometric time interval $\Delta q(t)$ passed from A to B for an integer observer $\cO_q$ (an ideal ranger using a $q$-clock in a plain world) would not be of 20 seconds:
\be\label{Det}
\De q(t) =\De t|1+\cT|\neq \De t\,.
\ee
Similarly, the spatial distance of the buoy from the beach was previously measured to be $\Delta x=|x_{\rm B}-x_{\rm A}|=100\,{\rm m}$, while the geometric distance would be
\be\label{Dex2}
\De q(x)=\De x|1+\cX|\neq \De x\,.
\ee
The speed of the sea turtle derived from the $t$-clock and the $x$-rod is
\be\label{speed}
v_x=v_q\left|\frac{1+\cT}{1+\cX}\right|\simeq v_q|1+\cT-\cX|\,,
\ee
where in the last step we assumed that $\cT,\cX\ll 1$ ($\De t\ll t_*$, $\De x\ll \ell_*$). Thus, the $x$-speed of the animal is $5\,{\rm m}\,{\rm s}^{-1}$ but the speed $v_q$ in the integer picture is different.

Clearly, if all experiments took place at the same spacetime and energy scales, the difference between the fractional and the integer picture would only be in the convention of the observer's measurement units. Living in a fractional world where a sea turtle takes 20\,s to reach a buoy 100\,m away would not be physically different from an integer world where the same event takes place in, say, 22 or 18\,s. Integer turtles would be slower or faster (albeit not tremendously so) than fractional turtles, but that would just be the normality for integer observers.

Suppose now that, at dawn, the same ranger observes a hatchling (of size $L'\sim 5\,{\rm cm}=O(10^{-1})\,L$ much greater than the characteristic length $\ell_*$) getting out from a nest in the sand, reaching the sea line unhindered by waves or curious tourists, entering the surf at the same point ${\rm A}'={\rm A}$ of the adult and reaching the same buoy B with a speed $v_{x'}=\De x'/\De t'$. Again, an integer observer $\cO_q$ would disagree with the measurements of the fractional observer $\cO_x$, but by a different relative amount because $\cT\neq \cT'$ and $\cX\neq \cX'$. 
 Then, using Eq.\ \Eq{speed}, one can determine the ratio of the speed of the adult over the speed of the hatchling in both frames:
\be\label{rr2}
r_x\simeq r_q |1+(\cT-\cT')+(\cX'-\cX)|\,,
\ee
where
\bs\ba
r_x&:=&\frac{v_x}{v_{x'}}=\frac{\Delta x/\Delta t}{\Delta x'/\Delta t'}\,,\\
r_q&:=&\frac{v_q}{v_{q'}}=\frac{\Delta q(x)/\Delta q(t)}{\Delta q(x')/\Delta q(t')}\,.
\ea\es

Before interpreting these results, we can indulge a bit more in our thought experiment and see what would happen if the fundamental scales were as large as $\ell_*\lesssim 1\,{\rm m}$ and $t_*\lesssim 10\,{\rm s}$. Then the corrections $\cT\lesssim 1$ and $\cX\lesssim 1$ would be comparable with, or even dominate over, the standard term in the measure. It is convenient to separate between two cases, a \emph{time-like multi-fractal} where $\cX=0$ and a \emph{space-like multi-fractal} where $\cT=0$. In the time-like case, one has $v_q=v_x/|1+\cT|\simeq v_x/|\cT|\ll v_x$ in the extreme regime $|\cT|\gg 1$ and
\be\label{vv3}
v_q< v_x\qquad (\cX=0)
\ee
in general, except when $-2\leq\cT\leq 0$. In the space-like case $\cT=0$, we get $v_q=v_x|1+\cX|\simeq v_x|\cX|\gg v_x$ in the extreme regime $|\cX|\gg 1$ and
\be\label{vv4}
v_q>v_x\qquad (\cT=0)
\ee
otherwise, except when $-2\leq\cX\leq 0$. 
 Multi-fractals along both spatial and time directions can display a more complex behavior.

 
\subsection{Initial-point presentation and dimensionless observables}

In the initial-point presentation, $\Delta q_+(t)= q_*(t_{\rm B}-t_{\rm A})$ and $\Delta q_+(x)= q_*(x_{\rm B}-x_{\rm A})$. According to Eqs.\ \Eq{XT}, \Eq{Det}, and \Eq{Dex2}, the geometric time interval $\Delta q_+(t)$ and the spatial distance $\De q_+(x)$ between A to B for an integer observer $\cO_q$ would be longer than for $\cO_x$ because $\a_0,\a>0$ and $\cT_+,\cX_+>0$:
\be
\De q_+(t) > \De t\,,\qquad \De q_+(x) > \De x\,.
\ee
Depending on which between the spatial and the temporal correction dominates, we will have $v_q< v_x$ (time-like multi-fractal) or $v_q> v_x$ (space-like multi-fractal):
\ba
&& v_q< v_x\qquad (\cX_+=0)\,,\label{vv1}\\
&& v_q> v_x\qquad (\cT_+=0)\,,\label{vv2}
\ea
just as in the general case \Eq{vv3}, \Eq{vv4}. As $x_{{\rm A}',{\rm B}'}=x_{{\rm A},{\rm B}}$, one has $\cX=\cX'$ and the spatial correction in Eq.\ \Eq{rr2} cancels exactly. For the sake of the argument, let $\Delta t'=10\Delta t$. Then, since $\cT'_+=\cT_+/10^{1-\a_0}<\cT_+$, the temporal correction in Eq.\ \Eq{rr2} is positive and one has $r_q < r_x=10$.

Here is the crucial point at last. The ranger is acquainted with the fact that the ratio $r_x=({\rm speed})_{\rm adult}/({\rm speed})_{\rm hatchling}$ of the swimming speed in the fractional picture depends on details such as the physiology of these animals, the temperature and time at the moment of hatching, the temperature of water, and so on, but that the average ratio reaches the universal, empirical value $\langle r_x \rangle\simeq 10$: in general, adults are about ten times faster than hatchlings.\footnote{Disclaimer: This is a fictional situation with no connection with real life and there is no such thing as a speed law for sea turtles. Nevertheless, the numbers given in the text are plausible and the ratio $r_x\sim 10$ is of the correct order of magnitude for various species of sea turtles and according to extant observations. Consult \cite{SaWy,Eck02} and references therein for some studies on sea-turtle speeds when swimming and walking.} Repeating their measurements every night and morning of the hatching season and finding a distribution of results definitely peaked at values close to $r_x=10$, the ranger agrees qualitatively with the turtle-speed law both as a fractional and as an integer observer, $r_x\simeq r_q$. However, imagine now that the relative experimental uncertainty reached by the ranger is better than in previous experiments, to the point where it is smaller than the correction $\cT'-\cT$ in \Eq{rr2}. Then they also find a \emph{systematic} discrepancy of the data points in the integer picture and a deviation of $r_q$ to values smaller than $10$. 
Then the observer must conclude that they are living in a fractional world.

Thus, even if geometric coordinates $q^\mu(x^\mu)$ have the same units as coordinates $x^\mu$, the existence of \emph{measurable dimensionless quantities} such as the ratio $r$ in our fictitious example allows an observer to recognize whether the underlying geometry is standard or multi-scale. This is an aspect of the relation between measurements and frame choice not covered in previous discussions \cite{fra7,frc11,frc13} and constitutes one of the main messages of the paper.

In practice, the outcome of experiments is less exciting, especially in macroscopic physics such as that describing turtle sea-faring. The magnitude of the corrections $\cX$ and $\cT$ are unknown since $\ell_*$ and $t_*$ are free parameters of the theory. Therefore, the error bar can at most place an upper bound on such correction and the experiment cannot distinguish between a fractional and an integer world. Unhappy with this situation, the wildlife ranger may decide to change clothes and go to some laboratory or particle accelerator to perform altogether different experiments, this time involving atoms and quantum particles. For instance, they might want to check out the spectral lines of light atoms or the relativistic quantum particles generated in scattering events at high energies. This is precisely the type of observations considered in \cite{frc12,frc14}, where upper bounds on $t_*$ and $\ell_*$ have been derived recently. These bounds are about 20--30 orders of magnitude smaller than the scales involved in the sea-turtle observation ($t_*<10^{-27}{\rm s}$, $\ell_*<10^{-19}{\rm m}$) and there would be no way to discriminate a turtle on such a multi-fractal from one on ordinary earth.

 
\subsection{Final-point presentation}\label{fpp}

Since $\cT_-=-\cT_+<0$ and $\cX_-=-\cX_+<0$, the situation for time-like and space-like fractals is reversed in the final-point presentation. Intervals \Eq{Det} and \Eq{Dex2} are shorter with respect to the fractional picture,
\be
\De q_-(t) < \De t\,,\qquad \De q_-(x) < \De x\,,
\ee
while the velocities obey
\ba
&& v_q> v_x\qquad (\cX_-=0)\,,\label{vv1-}\\
&& v_q< v_x\qquad (\cT_-=0)\,.\label{vv2-}
\ea
Also, $r_x< r_q$ in the turtle example and there is a systematic excess in the geometric ratio $r_q$ with respect to the observed value.

 
\subsection{Space- and time-like fractals}\label{3d}

We now pause for a moment and discuss a subtle point connecting some independent data we collected so far. One is that particles diffuse at a slower speed in a ``traditional'' space-like fractal, ${\ds}^{\rm space}<{\dh}^{\rm space}$. Another is that non-relativistic velocities are slower on a space-like multi-fractional geometry in the initial-point presentation and faster in the final-point presentation. Then, if we reasonably assume that the local non-relativistic velocities of a coarse-grained random motion\footnote{A nowhere differentiable curve does not admit local tangents and the argument in the text does not apply to an ideal random walk.} follow the same trend discovered for a macroscopic body, we can infer that spacetimes with $q$-derivatives are space-like fractals when the measure is in the initial-point presentation [Eq.\ \Eq{vv2}]. Moreover, if $\cT=0$ then one could represent a genuine diffusion process on space with Eq.~\Eq{pqb} where $q_\b(\s)=t$ and $\N_q^2$ is only spatial. However, this would not match with the discussion below Eq.\ \Eq{pqb} about the possibility of having an anomalous clock $\s$ when diffusing on traditional fractals. This observation does not place any strong constraint on the theory because we have seen that spacetime fractals do not have the same properties as traditional ones, but it leads us to consider as fractal also spacetimes with a time-like multi-scale geometry in the final-point configuration [Eq.\ \Eq{vv1-}]. However, we should also take into account that the general behavior \Eq{vv3}, \Eq{vv4} is reproduced without pathologies only in the initial-point presentation, while in the final-point presentation it is possible to hit the intervals 
\be\label{fibo}
-2<\cT<0\,,\qquad -2<\cX<0
\ee
and to reach the opposite regime \Eq{vv1-}, \Eq{vv2-}.

In reality, the three pairs of cases \Eq{vv3}--\Eq{vv4}, \Eq{vv1}--\Eq{vv2}, and \Eq{vv1-}--\Eq{vv2-} are an idealization of more complicated configurations. The main and most obvious reason is that space and time are entangled and, if the time direction is multi-scale, then the corrections $\cT$ and $\cX$ in Eq.\ \Eq{speed} compete with opposite signs and they can produce velocities $v_x=v_q|1-|\cT_-|+|\cX_-||<v_q$ even in the final-point presentation, provided $|\cT_-|>|\cX_-|$. The sign of the overall multi-scale correction in Eq.\ \Eq{speed} can be generic also in the null presentation \Eq{nullX}, depending on the details of the problem.

 
\section{Relativistic motion}\label{4}

In the theory with $q$-derivatives without gravity, the dynamics is invariant under the $q$-Poincaré transformations \Eq{qlort}, which are linear in the geometric coordinates but non-linear in the fractional coordinates. The components of the Lorentz matrices $\Lambda_\nu^{\ \mu}$ are standard and so is all the apparatus of special relativity when written in geometric coordinates. In particular, the Lorentz transformations of time and space are
\bs\ba
q^0(t')&=&\g\left[q^0(t)-\frac{v_q q^1(x^1)}{c^2}\right]\,,\\
q^1({x'}^1)&=&\g\left[q^1(x^1)-v_q q^0(t)\right]\,,\quad {x'}^2=x^2\,,\dots\,,
\ea\es
where the Lorentz factor $\g$ is \cite{frc10}
\be\label{gaq}
\g = \frac{1}{\sqrt{1-\frac{v_q^2}{c^2}}}
\ee
and $c$ is the geometric speed of light (i.e., the speed of light in the integer picture), which is constant. The line element and dynamics of a relativistic particle are discussed in \cite{frc10}, while the speed of light $c_x$ in the fractional picture is computed in a parallel work \cite{q-GW}. From the dispersion relation of photons in electrodynamics, it can be shown that the magnitude of the difference $\De c=c_x-c$ depends non-trivially on the energy of the photon and that $\De c=- O(1)\,(E/E_*)^{1-\a}<0$ for a space-like fractal spacetime (the actual speed of light is smaller than $c$) and $\De c=O(1)\,(E/E_*)^{1-\a_0}>0$ for a time-like fractal spacetime ($c_x>c$) \cite{q-GW}. Also, for a specific presentation of the measure in momentum space $\De c\equiv 0$ provided $\a_0=\a$. In any case, $\De c$ is experimentally constrained to be so small that it can be neglected here, $c_x\approx c$.

The Lorentz factor \Eq{gaq} in the integer picture yields a classic result of special relativity: the Galilean velocity $v_q$ of a body cannot exceed the geometric speed of light. However, from Eq.\ \Eq{speed} we have seen that $v_q= v_x/f$, where $f=|(1+\cT)/(1+\cX)|$ is the multi-scale correction computed above. Then the upper bound on $v_x$ is not $c$ but
\be
v_q= \frac{v_x}{f}< c\quad \Rightarrow\quad v_x< fc\,.
\ee
For a space-like fractal spacetime, $f=|1+\cX|^{-1}<1$ in the initial-point presentation for any $\cX>0$ or in the final-point presentation for $\cX<-2$. At microscopic scales, the speed limit is suppressed by a tiny factor $f$. This is consistent with both the sub-diffusion effect and the non-relativistic result.

For a time-like fractal spacetime, $f=|1+\cT|>1$ in the initial-point presentation for any $\cT>0$ or in the final-point presentation for $\cT<-2$. The Galilean speed $v_x$ in the fractional picture \emph{can} exceed the geometric speed of light by a factor $f$. For a macroscopic object this factor is mild and very close to 1, while for an object at scales $L\sim\ell_*$ and $t\sim t_*$ one can break the $c$-limit in a more spectacular way. Since $\ell_*$ and $t_*$ are at least as small as particle-physics scales, these results imply that one cannot use this multi-fractal theory to construct useful faster-than-light spacecraft.


\section{Future developments}\label{futu}

Our main results have already been outlined in Sect.\ \ref{intro} and we will not repeat them here. We rather comment upon two open subjects to be tackled in the future: the effect of logarithmic oscillations and the role of the presentation in the microscopic structure of multi-scale spacetimes.


\subsection{Log oscillations}

When log oscillations are taken into account, multi-fractional corrections become modulated and can change sign within the same time-like or space-like configuration. To maximize these effects, we assume that they all come from the oscillatory part of the leading term in the measure and that fractional power-law corrections are negligible. Assuming for simplicity one frequency and a vanishing amplitude for the sine contribution, for each spatial direction we have
\ba
q(x)&\simeq& x\left[1+ A\cos\left(\om\ln\frac{|x|}{\ell_\infty}\right)\right]\nonumber\\
&=:& x[1+F_\om(x)]\,,
\ea
plus a similar expression in time. Here $A>0$ and $\ell_\infty$ is a fundamental scale at the bottom of the whole hierarchy $\ell_\infty<\ell_1<\ell_2<\cdots$ of the most general measure. Given an interval $\De x=|x_{\rm B}-x_{\rm A}|$ and assuming $x_{\rm B}>x_{\rm A}$, the correction to the geometric distance can be approximated as
\ba
\De q(x)&=&\De x\left[1+F_\om(x_{\rm B})\right]\nonumber\\
&&+ x_{\rm A}\left[F_\om(x_{\rm B})-F_\om(x_{\rm A})\right]\,.\label{deqtot}
\ea
Naively, the easiest way to simplify this expression is to consider distances small enough to be sensitive to the fractal properties of the background. In this case, $\De x =\e \ell_\infty$, where $\e\ll 1$. Then $F_\om(x_{\rm B})\simeq F_\om(x_{\rm A})$ and the last term in Eq.\ \Eq{deqtot} cancels out. However, since we are on a fractal we can liberally apply the discrete scale invariance $x\to \la_\om^n\, x$ of the measure and note that $F_\om(x_{\rm B})\equiv F_\om(x_{\rm A})$ provided
\be\label{alax}
x_{\rm A}=\la_\om^n x_{\rm B}\,.
\ee
We can satisfy this condition by preparing the experiment. If $x_{\rm A}$ and $x_{\rm B}$ lie in the same copy of spacetime, then \Eq{alax} is automatic. If, by analogy with Cantor dusts, ``to lie in the same copy'' means to be in the same connected component at any given iteration (i.e., to belong to either $\cS_1(\cF)$ or $\cS_2(\cF)$), then by definition all causal experiments take place on the same copy and \Eq{alax} is guaranteed. If this tentative interpretation were not correct, one could fine tune the initial and final point by hand to obtain \Eq{alax}. Such a fine tuning is not strong; for instance, when $\a=1/2$ one has $\la_\om=1/N^2$ and the minimum tuning is of $1/4$.

In both cases, one gets (approximately or exactly)
\be
\De q(x)\simeq \De x\left[1+F_\om(x_{\rm B})\right]\,,\quad 1-A\leq\frac{\De q(x)}{\De x}\leq 1+A\,.
\ee
Regardless of the magnitude of the amplitude $A$, the correction $F_\om$ can take either sign.\footnote{If $A>1$, then there may even occur pathological situations where $\De q<0$. This happens because the measure is not positive definite for $A>1$ and it does not correspond, at ultra-microscopic scales, to an ordinary geometry. Still, it is a well-defined geometry, even if highly unconventional.}

In general, due to the modulation of the oscillations we would not be able to connect relations of the sort $v_x<v_q$ for non-relativistic velocities with the local velocities in a coarse-grained sub-diffusive stochastic process, even in a purely space-like or time-like fractal. The discussion of Sect.\ \ref{3d} would then need a revision. All these features will deserve to be explored in greater detail.


\subsection{Nowhere differentiability: toward stochastic spacetimes}\label{nodif}

Implicitly, in this paper we have begun to collect some evidence that there is a connection between the presentation choice and the stochastic properties of diffusion in these spacetimes. This interesting point went unnoticed in extant studies \cite{frc4,frc7} and it is worth looking into it in detail.

We have seen in Sect.\ \ref{3d} that, depending on the presentation of the measure, in certain time- or space-like systems the velocity of a non-relativistic body is slower than in an ordinary spacetime. Next, we have argued that this property is plausibly compatible with the microscopic sub-diffusion on fractals. The relation $v_x<v_q$ is valid outside the ``box'' \Eq{fibo} (i.e., the range of values for $\cT$ and $\cX$), which is not a region in the parameter space of the theory: both $\cT$ and $\cX$ depend on the measurements taken in the given system. If we require property C, then the only system-independent configurations are \Eq{vv1}, \Eq{vv2} and \Eq{vv1-}, \Eq{vv2-}. In that case, we set either the time or the spatial directions to be ordinary (this is part of the definition of the model) and the sign of the correction is unique for any system under consideration and for any regime. The box \Eq{fibo} should then be abandoned as a robust criterion for sub-diffusion.

However, we saw in Sect.\ \ref{2a} that property C holds independently of the presentation of the measure. The presentation, in fact, does not affect the scaling of $q$, nor any of the scaling relations which define the dimensions $\dh$, $\ds$, and $\dw$. The only element of ambiguity in these relations, fixed by an educated guess but never dispelled completely so far \cite{frc7}, was on the scaling $q_\b\sim \s^\b$ of the diffusion operator $\p/\p q_\b(\s)$ (and its multi-scale generalization) but not on its presentation. The universality of the spacetime dimension with respect to presentation choices is almost in accordance with the results \Eq{vv3}, \Eq{vv4}, valid everywhere in the parameter space except in the finite box \Eq{fibo}. Therefore, at least in the theory with $q$-derivatives, we must agree that:
\begin{itemize}
\item Spacetime space-like fractals have similar properties than ordinary spatial fractals, for instance the microscopic origin of sub-diffusion (i.e., the $v_x<v_q$ relation), but not in all presentations of the measure.
\item Diffusion and microscopic properties of spacetime time-like fractals or spacetime fractals with a non-trivial time-like component can differ widely from space-like ones without breaking the ABC rules.
\item Physical observables can provide elements to prefer one presentation over another but only in the case of a positive detection of multi-scale effects.
\item Although the dimensions of spacetimes are unaffected by the presentation of the measure, there is some yet unknown non-trivial relation between the presentation and microscopic stochastic properties of diffusion (see also Sect.\ \ref{2a}).
\end{itemize}

Having already analyzed the first three points, we can be more quantitative about the last. The choice of presentation of the integration measure discussed in Sect.\ \ref{2b} is intriguingly similar to the dilemma between the It\^{o} \cite{Ito44} and Stratonovich \cite{Str66} interpretation of stochastic integrals (see \cite{Oks03,Gar04} for extensive accounts). Given a stochastic process $X(t)$ (i.e., a sequence of random variables) and a Wiener process $B(t)$ (also known as standard Brownian motion), the solution of the stochastic differential equation $\dot X(t)=a[X(t),t]+f[X(t),t] \G(t)$ with initial condition $X(t_{\rm i})$ is
\ba
X(t)&=&X(t_{\rm i})+\int_{t_{\rm i}}^{t}\rmd t'\,a[X(t'),t']\nonumber\\
&&+\int_{t_{\rm i}}^t\rmd B(t')\,f[X(t'),t']\,,\label{stono}
\ea
where $\rmd B=\rmd t\,\G$. The first term is the initial point of the process. The second term drives the deterministic evolution of $X$ and poses no conceptual problem, contrary to the third term. The latter introduces a stochastic noise in such evolution and a source of ambiguity not apparent in the formal expression \Eq{stono}. In fact, $B(t)$ is nowhere differentiable,\footnote{Almost surely, i.e., with probability 1.} meaning that ordinary integer-order derivatives $\dot B$, $\ddot B$, \ldots are ill defined. Splitting the time interval $\De t=t-t_{\rm i}$ as $t_{\rm i}=t_0<t_1<\cdots< t_{n-1}=t$, without the pretense of being rigorous we can approximate the stochastic component in Eq.~\Eq{stono} as
\be\label{int0}
\int_{t_{\rm i}}^t\rmd B(t')\,f(t')=\lim_{n\to\infty}\sum_{j=0}^{n-1} f(\tilde t_j)\,[B(t_{j+1})-B(t_j)]\,.
\ee
For the Riemann--Stieltjes integral of an ordinary differentiable function $B(t)=b(t)$, the choice of the point $\tilde t_j\in [t_j,t_{j+1}]$ inside the interval $\De t_j=t_{j+1}-t_j$ is immaterial for the evaluation of the sum in the right-hand side: the result is unique and the Riemann sum is well defined. However, the Wiener process $B(t)$ fluctuates so much in $\De t_j$ that different choices of $\tilde t_j$ lead to inequivalent outcomes. 

In the It\^o interpretation, $\tilde t_j=t_j$ is the initial point in $\De t_j$ and the function $f$ only depends on the behavior of $B(t)$ up to the time $t_j$. In this case, the stochastic process $X(t)$ is a \emph{martingale}: the expectation value of an event at some future time $\tilde t_j\in (t_j,t_{j+1}]$ is equal to the value observed at the present time $t_j$. In other words, the knowledge of all previously observed values does not help to predict future outcomes.\footnote{For this reason, martingales are used as theoretical models of fair games.} In the Stratonovich interpretation, $\tilde t_j=(t_{j+1}+t_j)/2$ is taken in the middle of the interval and one symmetrizes between past and future (see the end of Sect.\ 4.3.6 of \cite{Gar04} for caveats). In this case, $X(t)$ is not a martingale: knowledge of prior outcomes may help to determine future events.

Both interpretations are valid but not in absolute terms: they describe systems with different stochastic properties (e.g., \cite{MoWe}). Similarly, both the initial- and the final-point presentations of the multi-fractional measure correspond to the same class of spacetimes but with different prescriptions on the volume of unit balls \cite{frc1}. Multi-fractional integrals are of a form similar to Eq.\ \Eq{int0}. Considering a class of measures $q(x-\bar x)$ all with the same scaling, we have
\ba
&&\int_{x_{\rm i}}^{x}\rmd q(x'-\bar x)\,f(x')=\int_{x_0}^{x-\bar x}\rmd q(x'')\,f(x''+\bar x)\nonumber\\
&&\qquad=\lim_{n\to\infty}\sum_{j=0}^{n-1} f(\tilde x_j+\bar x)[q(x_{j+1})-q(x_j)],\label{last}
\ea
where $x_0=x_{\rm i}-\bar x$. On a discontinuous genuine fractal, we would have a nowhere differentiable measure $q(x)$ (see, e.g., \cite{KoG} and the discussion in \cite{frc1}) and the same interpretation dilemma as in Eq.\ \Eq{int0}. In the theory with $q$-derivatives, on the other hand, $q(x)$ is differentiable and there is no ambiguity in $\tilde x_j\in [x_j,x_{j+1}]$. What we have, instead, is a presentation ambiguity which reflects in the choice of boundary conditions, i.e., the integration interval $[x_0,x_{n-1}]$. In Table \ref{tab1} we list the choices for the four different presentations discussed in the text.
\begin{table}[ht]
\begin{center}
\begin{tabular}{lccc}\hline
Presentation                           & $\bar x$ 							 & $x_0$      			 & $x_{n-1}$ \\\hline
Null							             				 & $0$         						 & $x_{\rm i}$			 & $x$       \\
Initial-point											     & $x_{\rm i}$						 & 0         		     & $\De x$   \\
Final-point							               & $x$ 										 & $-\De x$   		   & 0   			 \\
Symmetrized														 & $\frac{x_{\rm i}+x}{2}$ & $\frac{\De x}{2}$ & $-\frac{\De x}{2}$ \\\hline
\end{tabular}
\caption{Choice of boundaries in Eq.\ \Eq{last} for different presentations, where $\De x=x-x_{\rm i}$.\label{tab1}}
\end{center}
\end{table}

The change in the integration domain is more trivial than the ambiguity in the stochastic term \Eq{int0}. We can easily understand why. When we consider, for instance, volcano-like measure weights $\p_x q(x)\sim|x-\bar x|^{\a-1}$, we find an integrable singularity at $x=\bar x$. However, in a fractal we would expect to find a volcano at each and every point in the set, not just at a specific location $\bar x$. On a curved background, this issue is solved by general relativity: a local inertial frame centered on the observer is locally isomorphic to multi-scale Minkowski spacetime and each and every local inertial frame has its own volcano \cite{frc11}. Still, it may be desirable to have also a \emph{global} notion of irregularity, for instance when we work out particle-physics models on a flat Minkowski background. Such a global notion is not available in the theory with $q$-derivatives. A naive attempt to integrate over all possible $\bar x$ leads to ill-defined expressions of the form $\int_{-\infty}^{+\infty}\rmd\bar x\int_{-\infty}^{+\infty}\rmd x\,q(x-\bar x)\,\cL=\int_{-\infty}^{+\infty}\rmd x[\int_{-\infty}^{+\infty}\rmd \bar x\,q(x-\bar x)]\cL$, unless the ambient spacetime is compact. But even in a a compact space, this modification of the measure would not correspond to ``volcanoes everywhere.'' This is clear in the discretized version of the integrals:
\ba
&&\int_{x_{\rm i}}^{x}\rmd\bar x\int_{x_{\rm i}}^{x}\rmd x'\,q(x'-\bar x)\,f(x')\nonumber\\
&& \qquad=\int_{x_{\rm i}}^{x}\rmd\bar x\int_{x_0}^{x-\bar x}\rmd x''\,q(x'')\,f(x''+\bar x)\nonumber\\
&& \qquad=\lim_{n\to\infty}\sum_{k=0}^{n-1}\sum_{j=0}^{n-1} f(\tilde x_j+\bar x_k)\,[q(x_{j+1})-q(x_j)]\nonumber\\
&& \qquad\neq \lim_{n\to\infty}\sum_{j=0}^{n-1} f(\tilde x_j+\bar x_j)\,[q(x_{j+1})-q(x_j)]\,.
\ea
The last line is exactly what we would need: picking different presentations would amount to a non-trivial selection of the argument $\tilde x_j+\bar x_j$.

We should bear in mind that the theory with $q$-derivatives is a simplified version of a genuine nowhere differentiable fractal. Barring explicit and most challenging constructions of field theories on fractals \cite{Svo87,Ey89a,Ey89b} or field theories constructed with stochastic measures, the closest thing mimicking nowhere differentiability is fractional calculus. The multi-scale theory with fractional derivatives \cite{frc1,frc2} incarnates precisely this possibility and it may be the only multi-fractional theory obeying property D in the introduction. In fact, the differential operators in the theory with ordinary derivatives are the usual partial derivatives $\p_x$. In the theory with weighted derivatives, they are weighted version of the same operators, $\p_x\to (\p_x q)^{-1/2}\p_x[(\p_x q)^{1/2}\,\cdot\,]$. In the theory with $q$-derivatives, they are again of integer order but with a different weight distribution, $\p_x\to (\p_x q)^{-1}\p_x$. These modifications of $\p_x$ and the non-trivial integration measure give all these models an ``irregular'' geometry but in a rather simple-minded way. On the other hand, the derivatives of the fourth multi-fractional theory are non-local integro-differential operators, which are known to capture the properties of sets not differentiable in the ordinary sense (see \cite{frc1,KoG} and references therein). 

Therefore, the theory with fractional derivatives may well be the only one to describe a multi-fractal geometry in the strong sense. Due to its higher technical challenges, this framework has not been explored as extensively as the other multi-fractional theories but the preliminary analysis in \cite{frc1,frc2} and work in progress show enticing properties that include an exotic particle content and an improved perturbative renormalizability (absent in the other cases \cite{frc9}). The arguments presented in this section add fuel to our curiosity and strongly suggest that spacetimes with a fractional integro-differential structure would be intrinsically stochastic. We hope to report on that soon.


\section*{Acknowledgments} I thank D.\ Oriti for useful comments and G.\ Musser Jr.\ for stimulating me to think about questions \Eq{quest1a}--\Eq{quest2} critically. This work is under a Ram\'on y Cajal contract.



\begin{thebibliography}{99}
\bibitem{tH93}  G.\ 't Hooft, \tia{Dimensional reduction in quantum gravity} \procsin{Salamfestschrift}{A.\ Ali, J.\ Ellis, S.\ Randjbar-Daemi}{World Scientific}{Singapore}{1993}. \oarX{gr-qc/9310026}
\bibitem{Car09} S.\ Carlip, \tia{Spontaneous dimensional reduction in short-distance quantum gravity?} \doinn{10.1063/1.3284402}{AIP Conf.\ Proc.}{1196}{72}{2009}. \arX{0909.3329}
\bibitem{fra1}  G.\ Calcagni, \tia{Fractal universe and quantum gravity} \doinn{10.1103/PhysRevLett.104.251301}{Phys.\ Rev.\ Lett.}{104}{251301}{2010}. \arX{0912.3142}
\bibitem{AJL4}  J.\ Ambj{\o}rn, J.\ Jurkiewicz, R.\ Loll, \tia{Spectral dimension of the universe} \doinn{10.1103/PhysRevLett.95.171301}{Phys.\ Rev.\ Lett.}{95}{171301}{2005}. \oarX{hep-th/0505113}
\bibitem{BeH}   D.\ Benedetti, J.\ Henson, \tia{Spectral geometry as a probe of quantum spacetime} \doin{10.1103/PhysRevD.80.124036}{Phys.\ Rev.}{D}{80}{124036}{2009}. \arX{0911.0401}
\bibitem{SVW1}  T.P.\ Sotiriou, M.\ Visser, S.\ Weinfurtner, \tia{Spectral dimension as a probe of the ultraviolet continuum regime of causal dynamical triangulations} \doinn{10.1103/PhysRevLett.107.131303}{Phys.\ Rev.\ Lett.}{107}{131303}{2011}. \arX{1105.5646}
\bibitem{LaR5}  O.\ Lauscher, M.\ Reuter, \tia{Fractal spacetime structure in asymptotically safe gravity} \doij{10.1088/1126-6708/2005/10/050}{JHEP}{10}{050}{2005}. \oarX{hep-th/0508202}
\bibitem{CES}   G.\ Calcagni, A.\ Eichhorn, F.\ Saueressig, \tia{Probing the quantum nature of spacetime by diffusion} \doin{10.1103/PhysRevD.87.124028}{Phys.\ Rev.}{D}{87}{124028}{2013}. \arX{1304.7247}
\bibitem{Mod08} L.\ Modesto, \textit{Fractal structure of loop quantum gravity} \doinn{10.1088/0264-9381/26/24/242002}{Classical Quant.\ Grav.}{26}{242002}{2009}. \arX{0812.2214}
\bibitem{COT2}  G.\ Calcagni, D.\ Oriti, J.\ Th\"urigen, \tia{Spectral dimension of quantum geometries} \doinn{10.1088/0264-9381/31/13/135014}{Class.\ Quantum Grav.}{31}{135014}{2014}. \arX{1311.3340}
\bibitem{COT3}  G.\ Calcagni, D.\ Oriti, J.\ Th\"urigen, \tia{Dimensional flow in discrete quantum geometries} \doin{10.1103/PhysRevD.91.084047}{Phys.\ Rev.}{D}{91}{084047}{2015}. \arX{1412.8390}
\bibitem{Hor3}  P.\ Ho\v{r}ava, \tia{Spectral dimension of the universe in quantum gravity at a Lifshitz point} \doinn{10.1103/PhysRevLett.102.161301}{Phys.\ Rev.\ Lett.}{102}{161301}{2009}. \arX{0902.3657}
\bibitem{Con06} A.\ Connes, \tia{Noncommutative geometry and the standard model with neutrino mixing} \doij{10.1088/1126-6708/2006/11/081}{JHEP}{11}{081}{2006}. \oarX{hep-th/0608226}
\bibitem{CCM}   A.H.\ Chamseddine, A.\ Connes, M.\ Marcolli, \tia{Gravity and the standard model with neutrino mixing} \doinn{10.4310/ATMP.2007.v11.n6.a3}{Adv.\ Theor.\ Math.\ Phys.}{11}{991}{2007}. \oarX{hep-th/0610241}
\bibitem{AA}    E.\ Alesci, M.\ Arzano, \tia{Anomalous dimension in semiclassical gravity} \doin{10.1016/j.physletb.2011.12.026}{Phys.\ Lett.}{B}{707}{272}{2012}. \arX{1108.1507}
\bibitem{Ben08} D.\ Benedetti, \tia{Fractal properties of quantum spacetime} \doinn{10.1103/PhysRevLett.102.111303}{Phys.\ Rev.\ Lett.}{102}{111303}{2009}. \arX{0811.1396}
\bibitem{ACOS}  M.\ Arzano, G.\ Calcagni, D.\ Oriti, M.\ Scalisi, \tia{Fractional and noncommutative spacetimes} \doin{10.1103/PhysRevD.84.125002}{Phys.\ Rev.}{D}{84}{125002}{2011}. \arX{1107.5308}
\bibitem{ArTr1} M.\ Arzano, T.\ Trze\'sniewski, \tia{Diffusion on $\kappa$-Minkowski space} \doin{10.1103/PhysRevD.89.124024}{Phys.\ Rev.}{D}{89}{124024}{2014}. \arX{1404.4762}
\bibitem{Mod11} L.\ Modesto, \tia{Super-renormalizable quantum gravity} \doin{10.1103/PhysRevD.86.044005}{Phys.\ Rev.}{D}{86}{044005}{2012}. \arX{1107.2403}
\bibitem{CMNa}  G.\ Calcagni, L.\ Modesto, G.\ Nardelli, \tia{Quantum spectral dimension in quantum field theory} Int.\ J.\ Mod.\ Phys.\ D (2016, to appear). \arX{1408.0199}
\bibitem{CaG}   S.\ Carlip, D.\ Grumiller, \tia{Lower bound on the spectral dimension near a black hole} \doin{10.1103/PhysRevD.84.084029}{Phys.\ Rev.}{D}{84}{084029}{2011}. \arX{1108.4686}
\bibitem{Mur12} J.R.\ Mureika, \tia{Primordial black hole evaporation and spontaneous dimensional reduction} \doin{10.1016/j.physletb.2012.08.029}{Phys.\ Lett.}{B}{716}{171}{2012}. \arX{1204.3619}
\bibitem{AC1}   M.\ Arzano, G.\ Calcagni, \tia{Black-hole entropy and minimal diffusion} \doin{10.1103/PhysRevD.88.084017}{Phys.\ Rev.}{D}{88}{084017}{2013}. \arX{1307.6122}
\bibitem{MoN}   L.\ Modesto, P.\ Nicolini, \tia{Spectral dimension of a quantum universe} \doin{10.1103/PhysRevD.81.104040}{Phys.\ Rev.}{D}{81}{104040}{2010}. \arX{0912.0220}
\bibitem{DJW1}  B.\ Durhuus, T.\ Jonsson, J.F.\ Wheater, \tia{Random walks on combs} \doin{10.1088/0305-4470/39/5/002}{J.\ Phys.}{A}{39}{1009}{2006}. \oarX{hep-th/0509191}
\bibitem{AGW}   M.R.\ Atkin, G.\ Giasemidis, J.F.\ Wheater, \tia{Continuum random combs and scale dependent spectral dimension} \doin{10.1088/1751-8113/44/26/265001}{J.\ Phys.}{A}{44}{265001}{2011}. \arX{1101.4174}
\bibitem{GWZ1}  G.\ Giasemidis, J.F.\ Wheater, S.\ Zohren, \tia{Dynamical dimensional reduction in toy models of $4D$ causal quantum gravity} \doin{10.1103/PhysRevD.86.081503}{Phys.\ Rev.}{D}{86}{081503(R)}{2012}. \arX{1202.2710}
\bibitem{GWZ2}  G.\  Giasemidis, J.F.\ Wheater, S.\ Zohren, \tia{Multigraph models for causal quantum gravity and scale dependent spectral dimension} \doin{10.1088/1751-8113/45/35/355001}{J.\ Phys.}{A}{45}{355001}{2012}. \arX{1202.6322}
\bibitem{EiMi}  A.\ Eichhorn, S.\ Mizera, \tia{Spectral dimension in causal set quantum gravity} \doinn{10.1088/0264-9381/31/12/125007}{Class.\ Quantum Grav.}{31}{125007}{2014}. \arX{1311.2530}
\bibitem{CaMo1} G.\ Calcagni, L.\ Modesto, \tia{Nonlocality in string theory} \doin{10.1088/1751-8113/47/35/355402}{J.\ Phys.}{A}{47}{355402}{2014}. \arX{1310.4957}
\bibitem{Fal03} K.\ Falconer, \book{Fractal Geometry}{Wiley}{New York}{U.S.A.}{2003}
\bibitem{Str03} R.S.\ Strichartz, \tia{Fractafolds based on the Sierpi\'nski gasket and their spectra} \doinn{10.1090/S0002-9947-03-03171-4}{Trans.\ Am.\ Math.\ Soc.}{355}{4019}{2003}
\bibitem{fra4}  G.\ Calcagni, \tia{Discrete to continuum transition in multifractal spacetimes} \doin{10.1103/PhysRevD.84.061501}{Phys.\ Rev.}{D}{84}{061501(R)}{2011}. \arX{1106.0295}
\bibitem{frc1}  G\ Calcagni, \tia{Geometry of fractional spaces} \doinn{10.4310/ATMP.2012.v16.n2.a5}{Adv.\ Theor.\ Math.\ Phys.}{16}{549}{2012}. \arX{1106.5787}
\bibitem{frc2}  G.\ Calcagni, \tia{Geometry and field theory in multi-fractional spacetime} \doij{10.1007/JHEP01(2012)065}{JHEP}{01}{065}{2012}. \arX{1107.5041}
\bibitem{fra7}  G.\ Calcagni, \tia{Multifractional spacetimes, asymptotic safety and Ho\v{r}ava--Lifshitz gravity} \doin{10.1142/S0217751X13500929}{Int.\ J.\ Mod.\ Phys.}{A}{28}{1350092}{2013}. \arX{1209.4376}
\bibitem{frc7}  G.\ Calcagni, G.\ Nardelli, \tia{Spectral dimension and diffusion in multiscale spacetimes} \doin{10.1103/PhysRevD.88.124025}{Phys.\ Rev.}{D}{88}{124025}{2013}. \arX{1304.2709}
\bibitem{frc10} G.\ Calcagni, \tia{Relativistic particle in multiscale spacetimes} \doin{10.1103/PhysRevD.88.065005}{Phys.\ Rev.}{D}{88}{065005}{2013}. \arX{1306.5965}
\bibitem{frc11} G.\ Calcagni, \tia{Multi-scale gravity and cosmology} \doij{10.1088/1475-7516/2013/12/041}{JCAP}{12}{041}{2013}. \arX{1307.6382}
\bibitem{frc12} G.\ Calcagni, G.\ Nardelli, D.\ Rodr\'iguez-Fern\'andez, \tia{Particle-physics constraints on multifractal spacetimes} \doin{10.1103/PhysRevD.93.025005}{Phys.\ Rev.}{D}{93}{025005}{2016}. \arX{1512.02621}
\bibitem{frc13} G.\ Calcagni, G.\ Nardelli, D.\ Rodr\'iguez-Fern\'andez, \tia{Standard Model in multi-scale theories and observational constraints} \arX{1512.06858}
\bibitem{frc4}  G.\ Calcagni, \tia{Diffusion in multiscale spacetimes} \doin{10.1103/PhysRevE.87.012123}{Phys.\ Rev.}{E}{87}{012123}{2013}. \arX{1205.5046}
\bibitem{Alc94} M.\ Alcubierre, \tia{The warp drive: hyperfast travel within general relativity} \doinn{10.1088/0264-9381/11/5/001}{Class.\ Quantum Grav.}{11}{L73}{1994}. \oarX{gr-qc/0009013}
\bibitem{Eve96} A.E.\ Everett, \tia{Warp drive and causality} \doin{10.1103/PhysRevD.53.7365}{Phys.\ Rev.}{D}{53}{7365}{1996}
\bibitem{VBLi}  M.\ Visser, B.\ Bassett, S.\ Liberati, \tia{Superluminal censorship} \doinn{10.1016/S0920-5632(00)00782-9}{Nucl.\ Phys.\ Proc.\ Suppl.}{88}{267}{2000}. \oarX{gr-qc/9810026}
\bibitem{FLiB}  S.\ Finazzi, S.\ Liberati, C.\ Barcel\'o, \tia{Semiclassical instability of dynamical warp drives} \doin{10.1103/PhysRevD.79.124017}{Phys.\ Rev.}{D}{79}{124017}{2009}. \arX{0904.0141}
\bibitem{CFLPa} A.\ Coutant, S.\ Finazzi, S.\ Liberati, R.\ Parentani, \tia{Impossibility of superluminal travel in Lorentz violating theories} \doin{10.1103/PhysRevD.85.064020}{Phys.\ Rev.}{D}{85}{064020}{2012}. \arX{1111.4356}
\bibitem{fra6}  G.\ Calcagni, \tia{Diffusion in quantum geometry} \doin{10.1103/PhysRevD.86.044021}{Phys.\ Rev.}{D}{86}{044021}{2012}. \arX{1204.2550}
\bibitem{Akk2}  E.\ Akkermans, G.V.\ Dunne, A.\ Teplyaev, \tia{Thermodynamics of photons on fractals} \doinn{10.1103/PhysRevLett.105.230407}{Phys.\ Rev.\ Lett.}{105}{230407}{2010}. \arX{1010.1148}
\bibitem{Akk12} E.\ Akkermans, \tia{\href{http://www.ams.org/books/conm/601/11962/conm601-11962.pdf}{\cob Statistical mechanics and quantum fields on fractals}} \proc{Fractal Geometry and Dynamical Systems in Pure and Applied Mathematics II: Fractals in Applied Mathematics}{D.\ Carfi, M.L.\ Lapidus, E.P.J.\ Pearse, M.\ van Frankenhuijsen}{AMS}{Providence}{U.S.A.}{2013}. \arX{1210.6763}
\bibitem{AAGM3} G.\ Amelino-Camelia, M.\ Arzano, G.\ Gubitosi, J.~Magueijo, \tia{Dimensional reduction in momentum space and scale-invariant cosmological fluctuations} \doin{10.1103/PhysRevD.88.103524}{Phys.\ Rev.}{D}{88}{103524}{2013}. \arX{1309.3999}
\bibitem{DHLM}  M.\ Desbrun, A.N.\ Hirani, M.\ Leok, J.E.\ Marsden, \tia{Discrete exterior calculus} \oarX{math/0508341}
\bibitem{BeHi}  N.\ Bell, A.N.\ Hirani, \tia{PyDEC: software and algorithms for discretization of exterior calculus} \doinn{10.1145/2382585.2382588}{ACM Trans.\ Math.\ Softw.}{39}{3}{2012}. \arX{1103.3076}
\bibitem{COT1}  G.\ Calcagni, D.\ Oriti, J.\ Th\"urigen, \tia{Laplacians on discrete and quantum geometries} \doinn{10.1088/0264-9381/30/12/125006}{Class.\ Quantum Grav.}{30}{125006}{2013}. \arX{1208.0354}
\bibitem{RSc1}  M.\ Reuter, J.-M.\ Schwindt, \tia{A minimal length from the cutoff modes in asymptotically safe quantum gravity} \doij{10.1088/1126-6708/2006/01/070}{JHEP}{01}{070}{2006}. \oarX{hep-th/0511021}
\bibitem{rov07} C.\ Rovelli, \book{Quantum Gravity}{Cambridge University Press}{Cambridge}{U.K.}{2007}
\bibitem{Man67} B.\ Mandelbrot, \tia{How long is the coast of Britain? Statistical self-similarity and fractional dimension} \doinn{10.1126/science.156.3775.636}{Science}{156}{636}{1967}
\bibitem{Vas03} D.V.\ Vassilevich, \tia{Heat kernel expansion: user's manual} \doinn{10.1016/j.physrep.2003.09.002}{Phys.\ Rep.}{388}{279}{2003}. \arX{hep-th/0306138}
\bibitem{Sok12} I.M.\ Sokolov, \tia{Models of anomalous diffusion in crowded environments} \doinn{10.1039/C2SM25701G}{Soft Matter}{8}{9043}{2012}
\bibitem{MeN}   R.\ Metzler, T.F.\ Nonnenmacher, \tia{Fractional diffusion: exact representations of spectral functions} \doin{10.1088/0305-4470/30/4/011}{J.\ Phys.}{A}{30}{1089}{1997}
\bibitem{ReS11} M.\ Reuter, F.\ Saueressig, \tia{Fractal space-times under the microscope: a renormalization group view on Monte Carlo data} \doij{10.1007/JHEP12(2011)012}{JHEP}{1112}{012}{2011}. \arX{1110.5224}
\bibitem{frc5}  G.\ Calcagni, G.\ Nardelli, M.\ Scalisi, \tia{Quantum mechanics in fractional and other anomalous spacetimes} \doinn{10.1063/1.4757647}{J.\ Math.\ Phys.}{53}{102110}{2012}. \arX{1207.4473}
\bibitem{RLWQ}  F.-Y.\ Ren, J.-R.\ Liang, X.-T.\ Wang, W.-Y.\ Qiu, \tia{Integrals and derivatives on net fractals} \doinn{10.1016/S0960-0779(02)00211-4}{Chaos Solitons Fractals}{16}{107}{2003}
\bibitem{NLM}   R.R.\ Nigmatullin, A.\ Le M\'ehaut\'e, \tia{Is there geometrical/physical meaning of the fractional integral with complex exponent?} \doinn{10.1016/j.jnoncrysol.2005.05.035}{J.\ Non-Cryst.\ Solids}{351}{2888}{2005}
\bibitem{frc14} G.\ Calcagni, S.\ Kuroyanagi, S.\ Tsujikawa, \tia{Cosmic microwave background in a multi-scale spacetime} (2016, to appear)
\bibitem{frc8}  G.\ Calcagni, J.\ Magueijo, D.\ Rodr\'iguez-Fern\'andez, \tia{Varying electric charge in multiscale spacetimes} \doin{10.1103/PhysRevD.89.024021}{Phys.\ Rev.}{D}{89}{024021}{2014}. \arX{1305.3497}
\bibitem{SaWy}  M.\ Salmon, J.\ Wyneken, \tia{Orientation and swimming behavior of hatchling loggerhead turtles \emph{Caretta caretta} L.~during their offshore migration} \ndoinn{http://www.science.fau.edu/biology/faculty/Wyneken/DOC050817-004.pdf}{J.\ Exp.\ Mar.\ Biol.\ Ecol.}{109}{137}{1987}
\bibitem{Eck02} S.A.\ Eckert, \tia{Swim speed and movement patterns of gravid leatherback sea turtles (\emph{Dermochelys coriacea}) at St Croix, US Virgin Islands} \ndoinn{http://jeb.biologists.org/content/205/23/3689}{J.\ Exp.\ Biol.}{205}{3689}{2002}
\bibitem{q-GW}  G.\ Calcagni, \tia{Lorentz violations in multifractal spacetimes} \arX{1603.03046}
\bibitem{Ito44} K.\ It\^o, \tia{Stochastic integral} \doinn{10.3792/pia/1195572786}{Proc.\ Imperial Acad.\ Tokyo}{20}{519}{1944}
\bibitem{Str66} R.L.\ Stratonovich, \tia{A new representation for stochastic integrals and equations} \doinn{10.1137/0304028}{SIAM J.\ Control}{4}{362}{1966}
\bibitem{Oks03} B.K.\ {\O}ksendal, \book{Stochastic Differential Equations: An Introduction with Applications}{Springer-Verlag}{Berlin}{Germany}{2003}
\bibitem{Gar04} C.W.\ Gardiner, \book{Handbook of Stochastic Methods}{Springer-Verlag}{Berlin}{Germany}{2004}
\bibitem{MoWe}  W.\ Moon, J.S.\ Wettlaufer, \tia{On the interpretation of Stratonovich calculus} \doinn{10.1088/1367-2630/16/5/055017}{New J.\ Phys.}{16}{055017}{2014}. \arX{1402.6895}
\bibitem{KoG}   K.M.\ Kolwankar, A.D.\ Gangal, \tia{Fractional differentiability of nowhere differentiable functions and dimensions} \doinn{10.1063/1.166197}{Chaos}{6}{505}{1996}. \oarX{chao-dyn/9609016}
\bibitem{Svo87} K.\ Svozil, \tia{Quantum field theory on fractal space-time} \doin{10.1088/0305-4470/20/12/033}{J.\ Phys.}{A}{20}{3861}{1987}
\bibitem{Ey89a} G.\ Eyink, \tia{Quantum field-theory models on fractal spacetime. I: Introduction and overview} \doinn{10.1007/BF01228344}{Commun.\ Math.\ Phys.}{125}{613}{1989}
\bibitem{Ey89b} G.\ Eyink, \tia{Quantum field-theory models on fractal spacetime. II: Hierarchical propagators} \doinn{10.1007/BF02124332}{Commun.\ Math.\ Phys.}{126}{85}{1989}
\bibitem{frc9}  G.\ Calcagni, G.\ Nardelli, \tia{Quantum field theory with varying couplings} \doin{10.1142/S0217751X14500122}{Int.\ J.\ Mod.\ Phys.}{A}{29}{1450012}{2014}. \arX{1306.0629}
\end{thebibliography}
\end{document}